\documentclass[12pt]{article}

\newenvironment{wileykeywords}{\textsf{Keywords:}\hspace{\stretch{1}}}{\hspace{\stretch{1}}\rule{1ex}{1ex}}

\usepackage{amsmath,amssymb}
\usepackage{graphicx}
\usepackage{color}
\usepackage{dcolumn}
\usepackage{bm}
\usepackage[numbers,super,comma,sort&compress]{natbib}

\usepackage{caption}
\usepackage{subfig}
\newcommand{\tn}{\textnormal}

\definecolor{background-color}{gray}{0.98}

\title{Utility of potential energy span as an approximate free energy proxy}
\author{Kai Wang\thanks{School of Life Sciences, Jilin University}, Lanru Liu\thanks{School of Life Sciences, Jilin University}, Pu Tian \thanks{School of Life Sciences and MOE Key Laboratory of Molecular Enzymology and Engineering, Jilin University, 2699 Qianjin Street, Changchun China 130012} }

\begin{document}

\maketitle

\begin{abstract}
Free energy calculation is critical in predictive tasks such as protein folding, docking and design. However, rigorous calculation of free energy change is prohibitively expensive in these practical applications. The minimum potential energy is therefore widely utilized to approximate free energy. In this study, based on analysis of extensive molecular dynamics (MD) simulation trajectories of a few native globular proteins, we found that change of minimum and corresponding maximum potential energy terms exhibit similar level of correlation with change of free energy. More importantly, we demonstrated that change of span (maximum - minimum) of potential energy terms, which engender negligible additional computational cost, exhibit considerably stronger correlations with change of free energy than the corresponding change of minimum and maximum potential energy terms. Therefore, potential energy span may serve as an alternative efficient approximate free energy proxy.
\end{abstract}

\begin{wileykeywords}
free energy, correlation, potential energy span, minimum potential energy approximation
\end{wileykeywords}

\clearpage


\begin{figure}[h]
\centering
\colorbox{background-color}{
\fbox{
\begin{minipage}{1.0\textwidth}
\includegraphics[height=40mm]{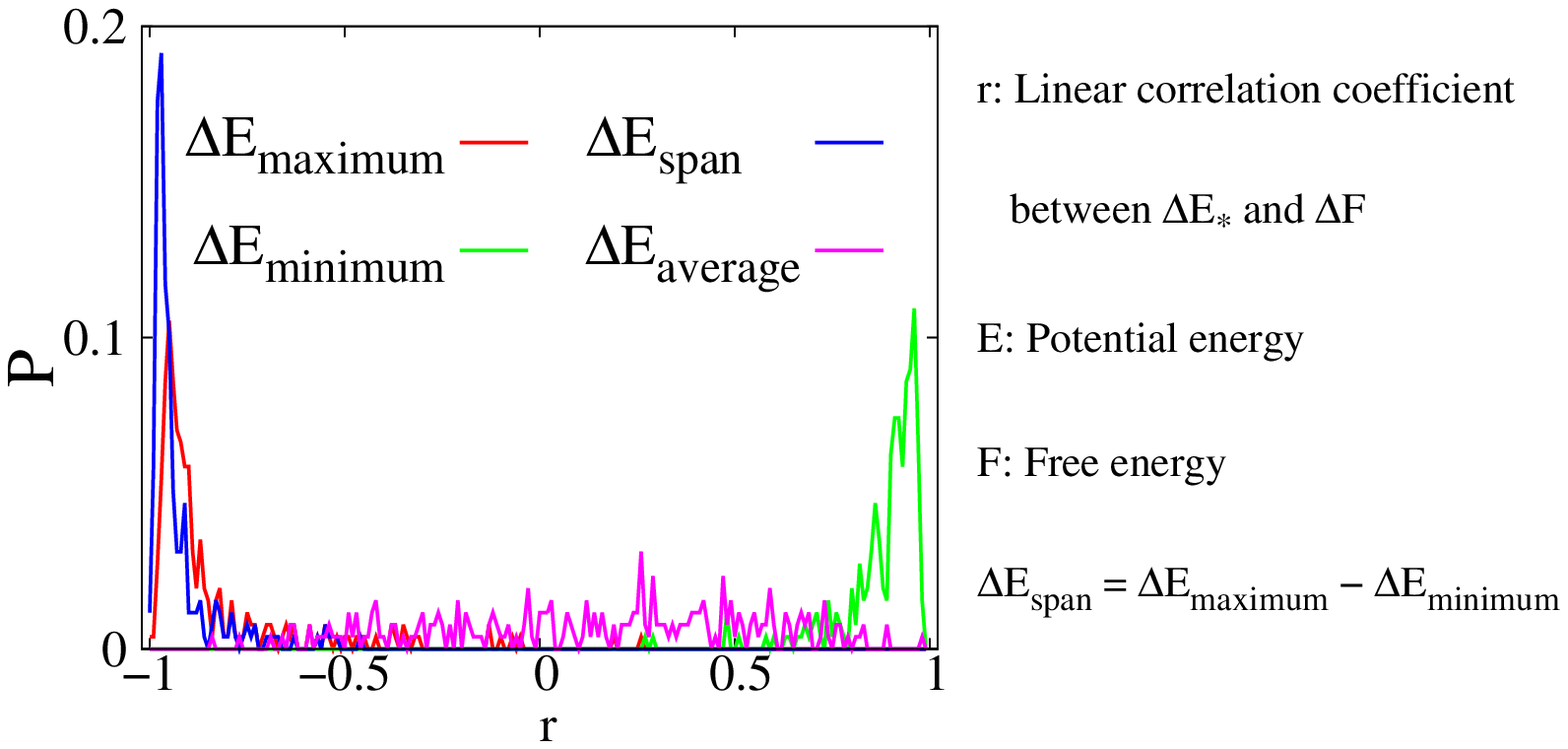} 
\\
Minimum potential energy is widely utilized to approximate free energy. We found that change of maximum potential energy from canonical distributions exhibit similar level of correlation with change of free energy to that observed between change of minimum potential energy and change of free energy. More importantly, change of span of potential energy consistently and significantly correlate better with free energy than change of both minimum and maximum potential energy. 
\end{minipage}
}}
\end{figure}

  \makeatletter
  \renewcommand\@biblabel[1]{#1.}
  \makeatother

\bibliographystyle{apsrev}

\renewcommand{\baselinestretch}{1.5}
\normalsize

\clearpage

\section*{\sffamily \Large INTRODUCTION} 
Estimating free energy changes for macrostate pairs is essential in protein folding, docking and design. In such tasks, one usually need to sample a large number of backbone conformational macrostates and identify those with the lowest free energy. Rigorous free energy calculation methods, such as those based on thermodynamic integration (TI)\cite{Kirkwood1935,Darve2001}, free energy perturbation (FEP)\cite{Zwanzig1954,Zwanzig1955,Bash1987} and non-equilibrium work (NEW)\cite{Jarzynski1997,Hummer2001,Goette2008}, are not suitable due to prohibitive computational cost\footnote{There are two major aspects for the cost. Firstly, each single calculation of free energy change between two macrostates is expensive; Secondly, to evaluate $n$ macrostates, these methods requires $\sim n^2$ calculations. Suppose we have $n$ macrostates $\{MS_1, MS_2, MS_3, \cdots, MS_n \}$ to be evaluated for their relative free energy via TI type of methods. If we calculate $\Delta F^{12}$ (the free energy change when the system goes from macrostates $MS_1$ to $MS_2$), $\Delta F^{23}$, $\cdots$, $\Delta F^{(n-1)n}$, and express $\Delta F^{1n}$ as the sum: $\Delta F^{12} + \Delta F^{23} + \cdots + \Delta F^{(n-1)n}$, we would have accumulation of error following a one-dimensional random walk, which is not bounded. To prevent such unbounded propagation of error, pairwise calculations are necessary and the resulting number of calculations increases from $\mathcal{O}(n)$ to $\mathcal{O}(n^2)$!}. As a necessary compromise, two types of approximate methodologies, including a large number of scoring functions\cite{Huang2010} and physics based methods such as MM/P(G)BSA\cite{MMPBSA-JMC} and linear interaction energy (LIE)\cite{LIE1998} model, have been developed and widely utilized. Despite its apparent caveat of neglecting entropic contributions, the minimum potential energy (MiPE) approximation is popular for its great simplicity\cite{Kamisetty2011}. When sampling is carried out for a given macrostate, significantly more information is produced in addition to MiPE. We are interested in extracting more information to improve the MiPE approximation while retaining its attractive simplicity. To avoid entanglement of errors caused by extent of sampling, quality of force fields and experimental measurements, direct comparison between computation and experiments is avoided. Instead, we utilized sufficiently well-converged molecular dynamics (MD) simulation trajectories, where arbitrary macrostates may be constructed through partition of visited configurational space, to evaluate quality of the MiPE approximation and explore possible improvement.   

Based on the analysis of MD trajectories of hen egg white lysozyme (HEWL) generated in canonical ensemble (see \emph{Methodology} for details), we found that both change of MiPE and maximum potential energy (MaPE) terms (hereafter denoted as $\Delta E^*_{min}$ and $\Delta E^*_{max}$, `*' is a wild card for ``$p$''-- protein self energy, ``$p\tn{-}sv$''--protein solvent interaction energy, ``$ppsv$''-- the sum of the previous two terms and ``$tot$''--the total potential energy of the simulated molecular system) have similar level of overall correlations with change of free energy ($\Delta F$) for corresponding macrostate pairs. 
More importantly, we found that change of span of potential energy terms ($\Delta E^*_{span} = \Delta (E^*_{max} - E^*_{min})$) correlate consistently and significantly better with change of free energy than change of both MiPE and MaPE terms. Analysis of MD trajectory sets for two other globular proteins with different folds and sizes revealed similar trends. Therefore, we think it is likely that span of potential energy terms are better free energy proxies than MiPE terms in globular proteins. Since calculating $E_{span}$ beyond MiPE engenders only an additional counter for MaPE and a subtraction operation, an essentially negligible cost in both memory and wall-clock time, we expect $E_{span}$ to be a useful free energy proxy in high throughput estimation of free energy differences in proteins, and possibly other complex biomolecular systems.

\section*{\sffamily \Large METHODOLOGY}
HEWL trajectory set is based on the $200\mu s$ trajectory set reported previously\cite{Wang2015}, $20,000$ snapshots were taken from which to seed short trajectories, each $10$-$ps$ long and with $1000$ snapshots recorded. The BPTI trajectory as reported in another study\cite{Wenzhao149} was extended to $\sim3.3\mu s$, 6124 uniformly distributed snapshots were taken to seed short trajectories, each $10$-$s$ long with 1000 snapshots recorded. The human sulfotransferase trajectory set was based on the $2A1+LCA$ system in another study\cite{Zhao2015a}, 3760 snapshots uniformly distributed in the original trajectory set (collectively $\sim 1.8\mu s$) were taken to seed $10$-$ps$ short trajectories, each has $1000$ snapshots recorded. Potential energy terms were calculated with the pairInteraction option of NAMD\cite{NAMD}. Macrostates were constructed by projection onto backbone dihedrals. Specifically, for a given dihedral has an observed range $[a, b]$, then 20 $\frac{b-a}{20}$ sized  bins are constructed. Alternatively, if a given dihedral has an observed range $[0, a]$ and $[b, 360]$, then 20 $\frac{360+a-b}{20}$ sized bins are constructed.



\section*{\sffamily \Large RESULTS}


\section*{\sffamily \Large Correlations between change of MiPE terms and change of free energy for HEWL}
 We first need to construct macrostates for evaluating performance of MiPE approximation. To this end, each backbone was utilized as an order parameter for projection and divided into 20 equally sized windows (see \emph{Methodology} for details), each of which is defined as a macrostate. Consequently there are 20 mutually exclusive and exhaustive macrostates, and 190 macrostate pairs associated with each backbone. 

In a set of sufficiently well converged MD trajectories, change of free energy upon transition from macrostates $A$ to $B$ may be calculated as:
\begin{equation}
\Delta F^{AB} = k_BTln\frac{N^A_{snap}}{N^B_{snap}}\label{eq:snap}
\end{equation} 
with $N^{A(B)}_{snap}$ being observed number of snapshots in macrostate $A(B)$, $k_B$ being Boltzmann constant and $T$ being the temperature. For each backbone dihedral, $\Delta F$ corresponds to each of the 190 associated macrostate pairs was calculated with equation \ref{eq:snap}. We also calculated corresponding change of MiPE terms, including $\Delta E^p_{min}$, $\Delta E^{p\tn{-}sv}_{min}$, $\Delta E^{ppsv}_{min}$ and $\Delta E^{tot}_{min}$. Linear correlations between changes of these MiPE terms ($\Delta E^*_{min}$) and corresponding changes of free energy ($\Delta F$) were analyzed. Such operation was repeated for all backbone dihedrals, each of which corresponds to a unique way of macrostates definition (configurational space partition). A summary of the results is presented in Fig. \ref{fig:EminFE}. 

Different change of MiPE terms exhibit comparable level of overall correlation with change of free energy. The consistency among different MiPE terms suggest strong correlations between protein self-energy, protein-solvent interaction energy and solvent energy, in agreement with our earlier analysis of correlations between these energetic terms\cite{Zhao2013}. Nevertheless, strong correlations are not equivalence, and slight differences exist among examined change of MiPE terms. To better characterize such differences, the cumulative probability density (CPD) of the correlation coefficients associated with individual macrostate-defining backbone dihedrals are plotted with respect to their absolute values and shown in Fig \ref{fig:CPDmmsa}a. Area under each of these curves indicate overall correlation between corresponding $\Delta E_{min}$ term and $\Delta F$. $\Delta E^{tot}_{min}$ exhibit the strongest correlation with $\Delta F$. For the remaining three change of MiPE terms, $\Delta E^{p}_{min}$ perform slightly better than $\Delta E^{ppsv}_{min}$ and $E^{p\tn{-}sv}_{min}$. Additionally, macrostates defined by different backbone dihedrals were pooled together (with 13104640 macrostate pairs in total) to perform an overall linear fit, qualitatively similar relative correlation with $\Delta F$ was observed for these change of MiPE terms ( see Fig. S2a, b, c, d).
Secondly, while consistently good correlations between change of MiPE terms and change of free energy are observed when macrostates are defined by projection onto backbone dihedral angles in stable secondary structures and by projection onto some of backbone dihedrals in loop regions, significantly weaker correlations are observed when macrostates are defined by projection onto the remaining backbone dihedral angles in flexible loop regions.

\section*{\sffamily \Large Correlations between change of MaPE terms and change of free energy for HEWL}
\begin{figure}
\centering 
\subfloat[]{\includegraphics[width=1.8in]{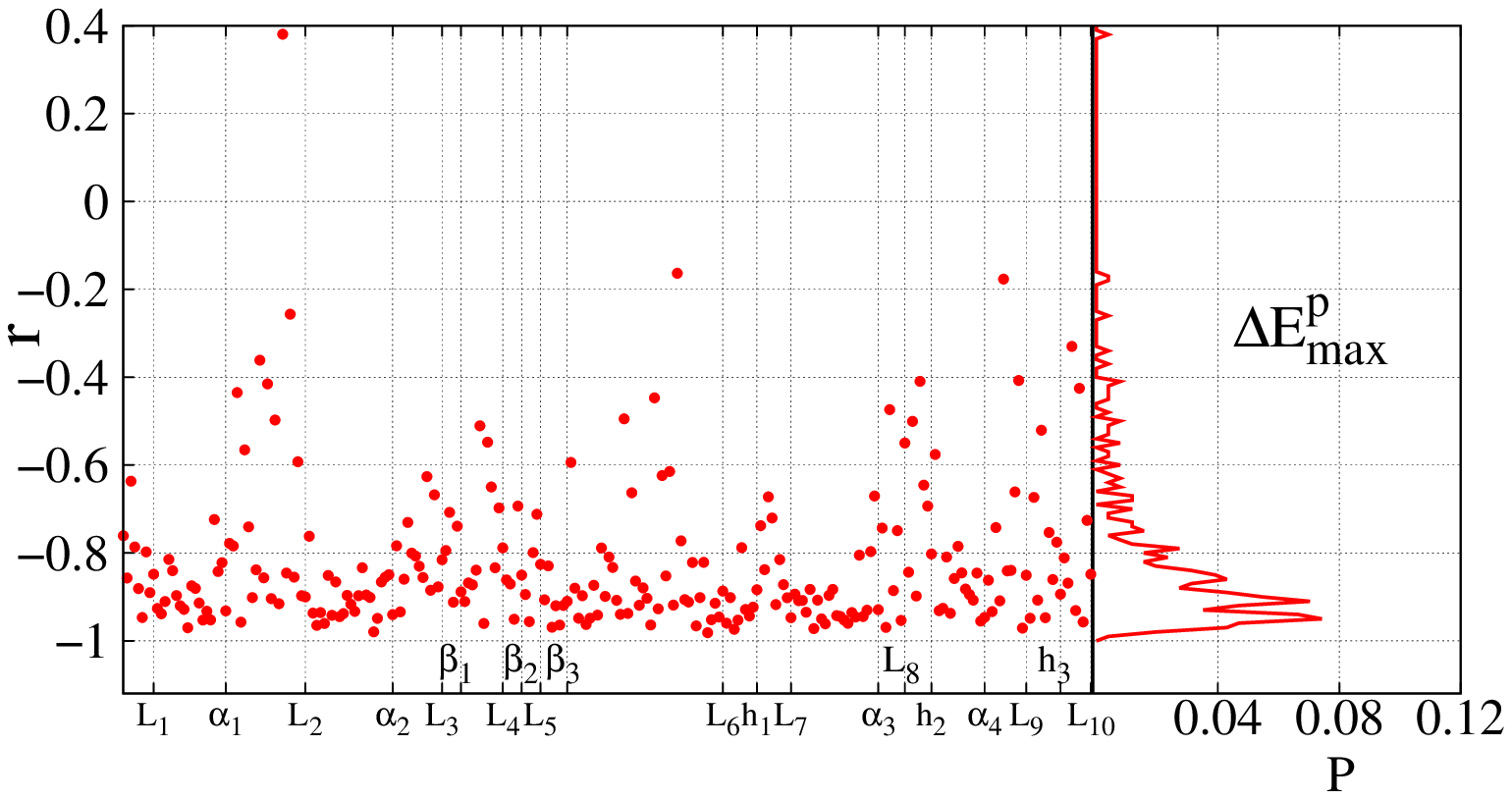}}
\subfloat[]{\includegraphics[width=1.8in]{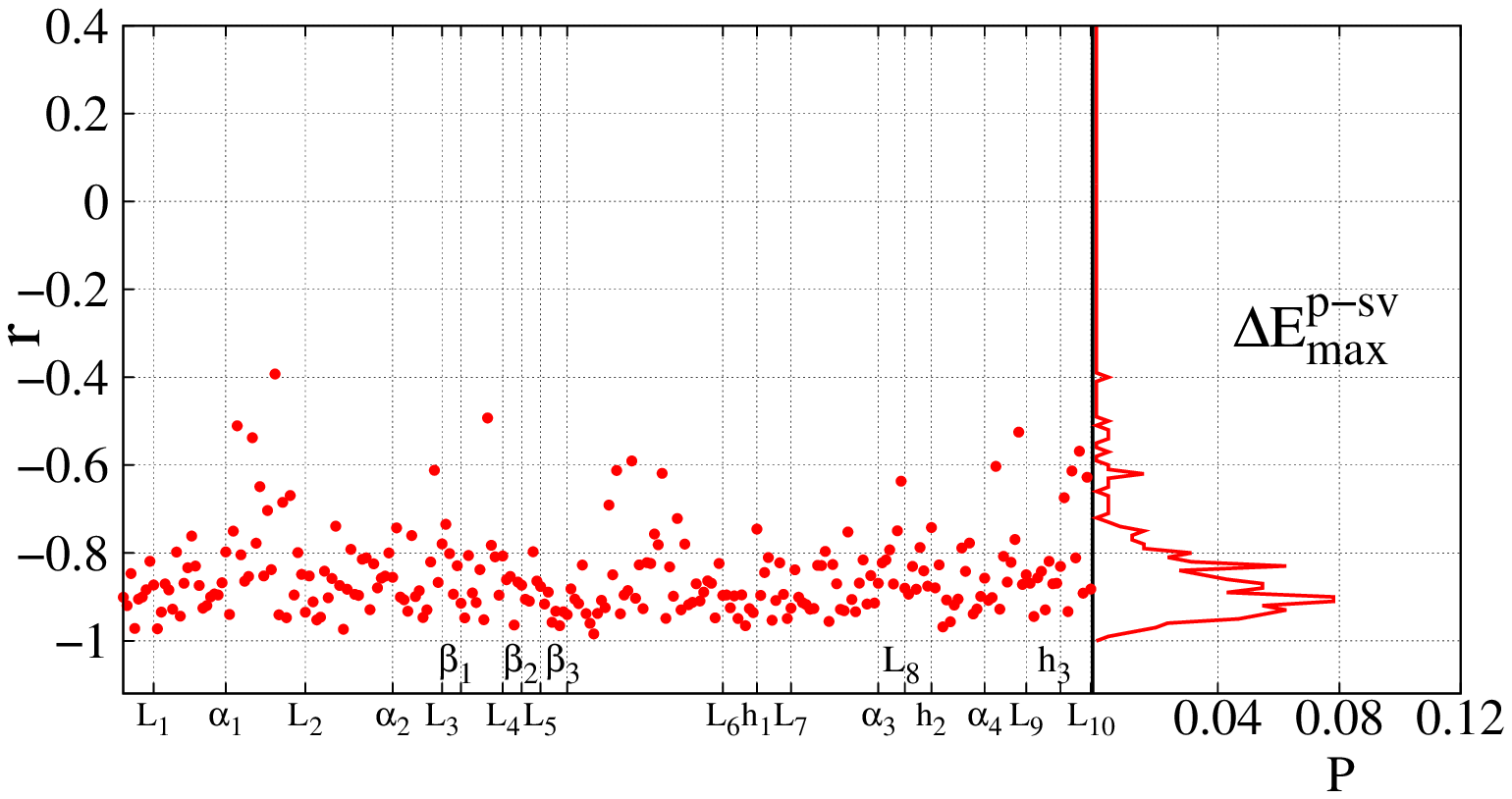}}\\
\subfloat[]{\includegraphics[width=1.8in]{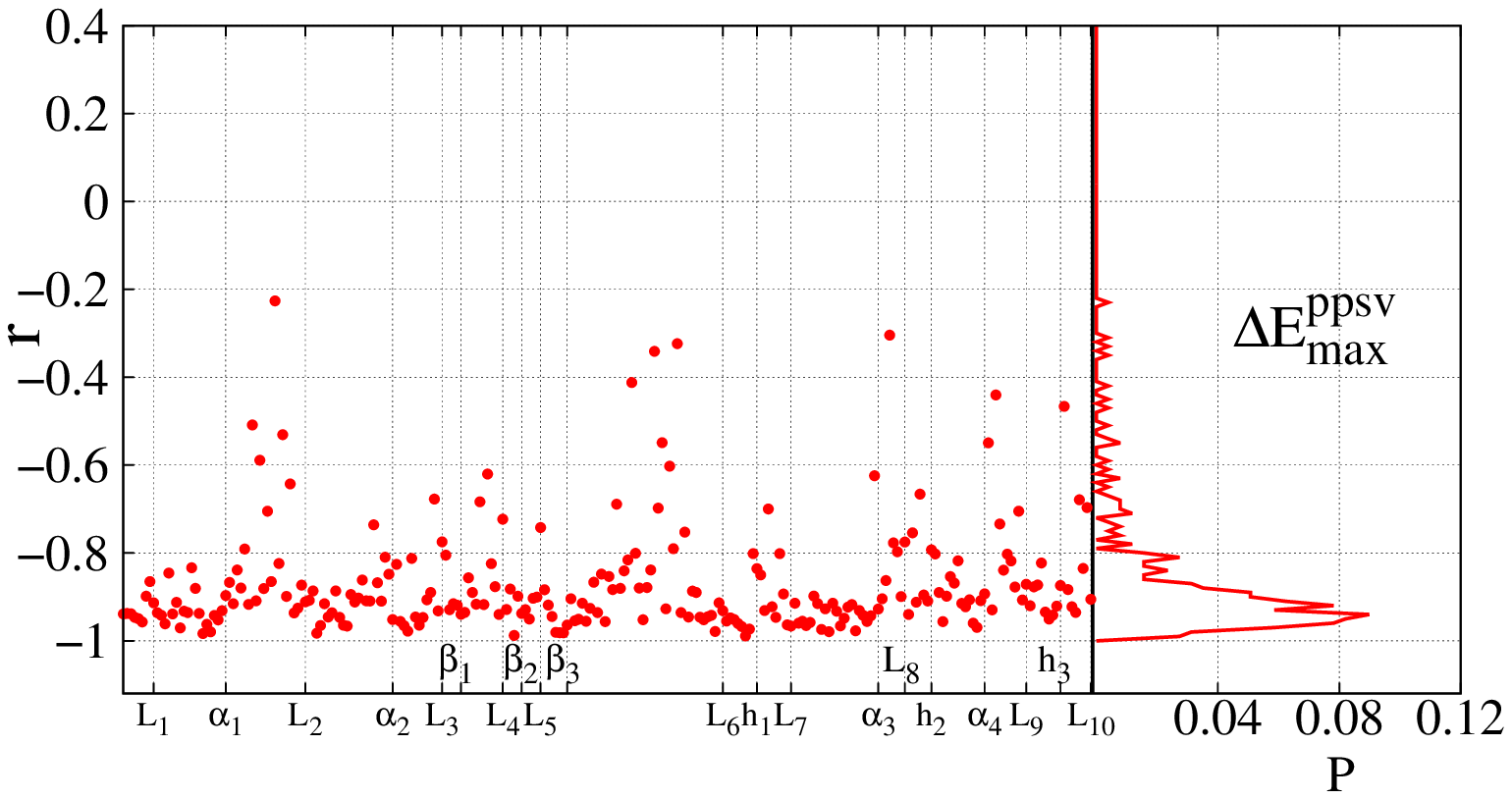}}
\subfloat[]{\includegraphics[width=1.8in]{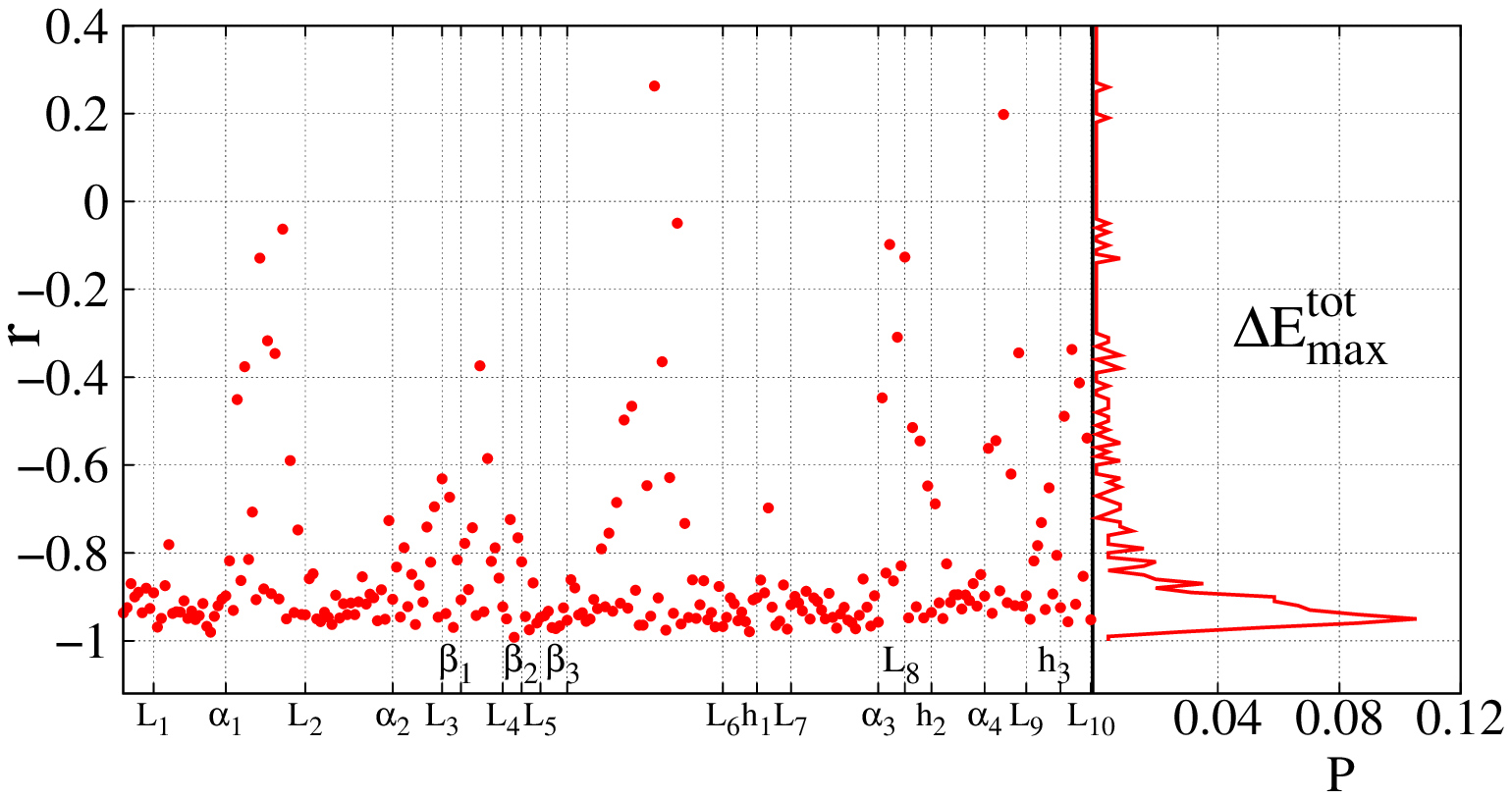}}\\
\caption{Linear correlations between $\Delta E^*_{max}$ (a $E^p_{max}$, b $E^{p\tn{-}sv}_{max}$, c $E^{ppsv}_{max}$ and d $E^{tot}_{max}$) and $\Delta F$. Left panels are scatter plots of linear correlation coefficients ($r$) between $\Delta E^*_{max}$  and $\Delta F$ for macrostates associated with each backbone dihedral, the corresponding secondary structures of backbone dihedrals are indicated on the horizontal axis. Right panels are the probability distributions of linear correlation coefficients observed in the 256 different ways of macrostates definition corresponding to 256 backbone dihedrals.}  
\label{fig:EmaxFE}
\end{figure}

Boltzmann distribution states that microstates with higher potential energy have smaller statistical weight (probability being observed). For macrostates with relatively higher observed potential energy, an alternative perspective would be that there are more high energy microstates in such macrostates. Therefore, higher observed potential energy qualitatively indicates larger entropy in the corresponding potential energy range. Prompted by this thought, we suspect that MaPE terms from canonical distributions might correlate with free energy, especially when entropy in the relatively higher potential energy range is important. We therefore analyzed correlations between change of MaPE terms with change of free energy in a similar way. A summary of the results was shown in Fig. \ref{fig:EmaxFE}. 
Indeed, consistently good correlations between $\Delta E^*_{max}$ and $\Delta F$ are observed when macrostates are defined by projection onto most of backbone dihedral angles. Significantly weaker correlations are observed for the same set of macrostate pairs where weaker correlations between $\Delta E^*_{min}$ and $\Delta F$ are observed. As one would intuitively expect, all $\Delta E^*_{max}$ terms negatively correlate with $\Delta F$. The order of correlation strength among $\Delta E^*_{max}$ terms and $\Delta F$ is slightly different from that observed for $\Delta E^*_{min}$ (see Fig. \ref{fig:CPDmmsa}a) and b)), with $\Delta E^{ppsv}_{max}$ exhibiting the strongest correlation with $\Delta F$. Similar to analysis of $\Delta E^*_{min}$, macrostates defined by different backbone dihedrals were pooled together (with 13104640 macrostate pairs in total) to perform an overall linear fit, the results were shown in the supporting Fig. S\ref{fig:EmaxFE}.  
 
\section*{\sffamily \Large Correlations between change of span of potential energy terms and change of free energy for HEWL}
\begin{figure}
\centering 
\subfloat[]{\includegraphics[width=1.8in]{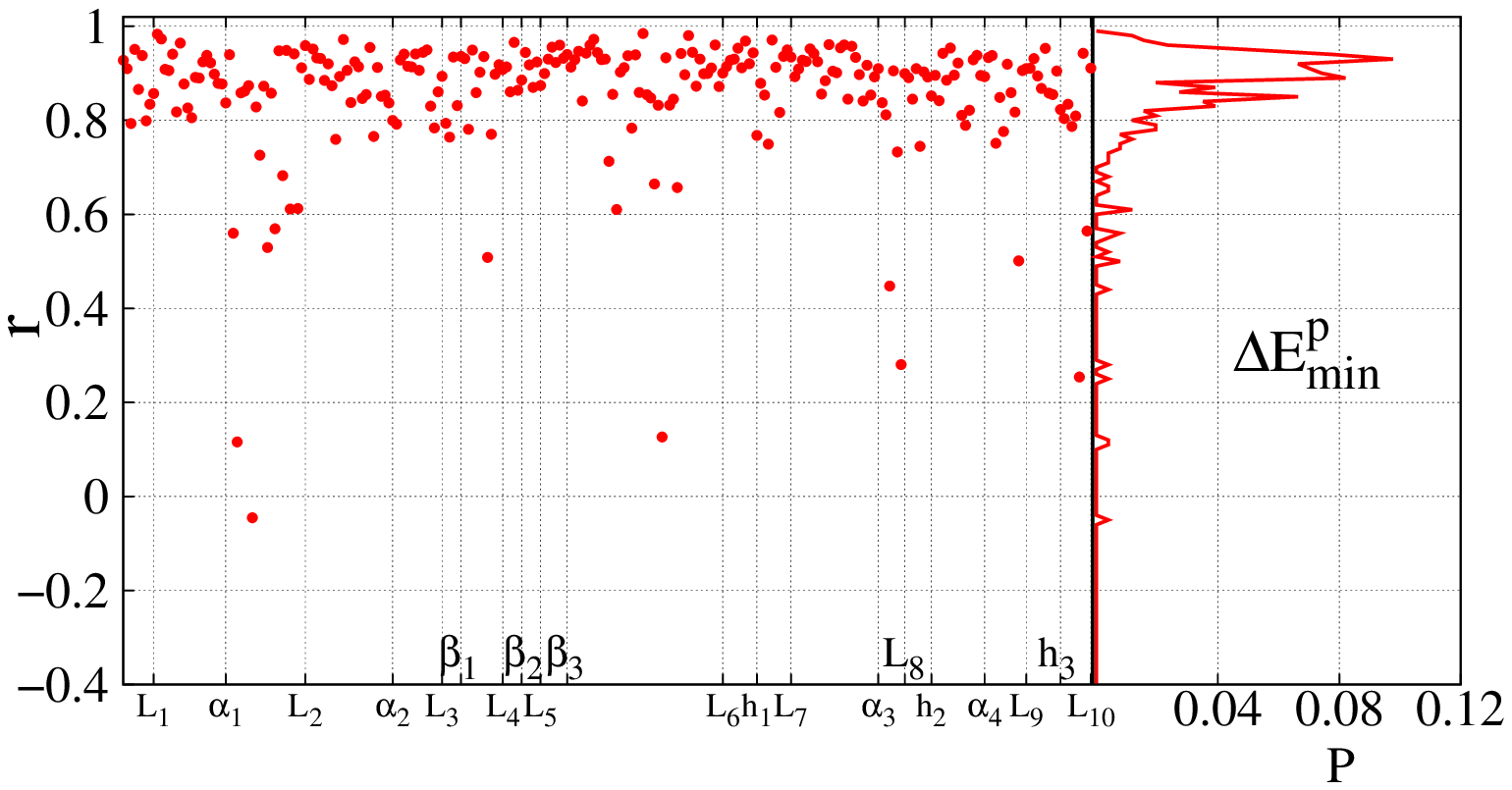}}
\subfloat[]{\includegraphics[width=1.8in]{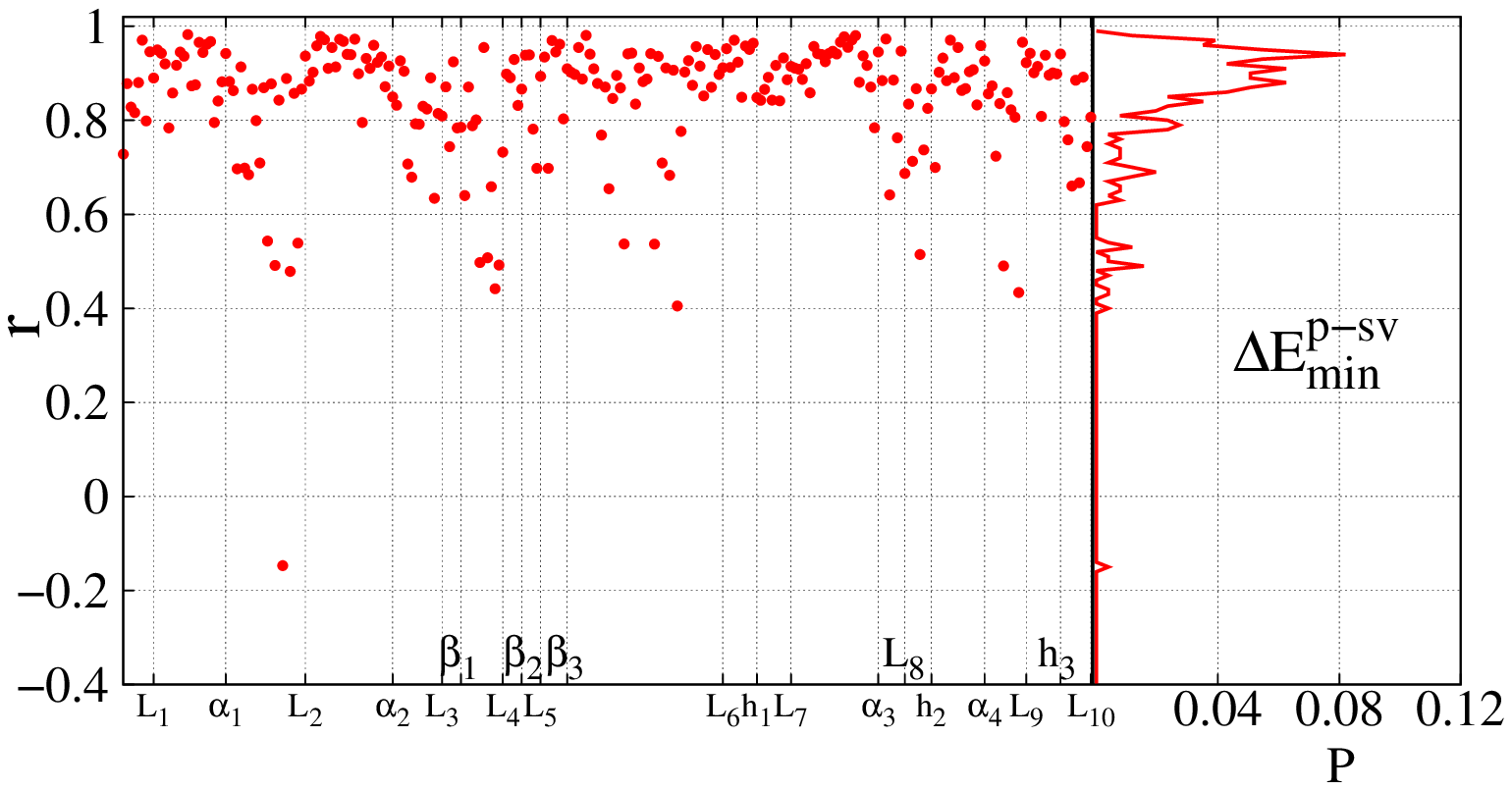}}\\
\subfloat[]{\includegraphics[width=1.8in]{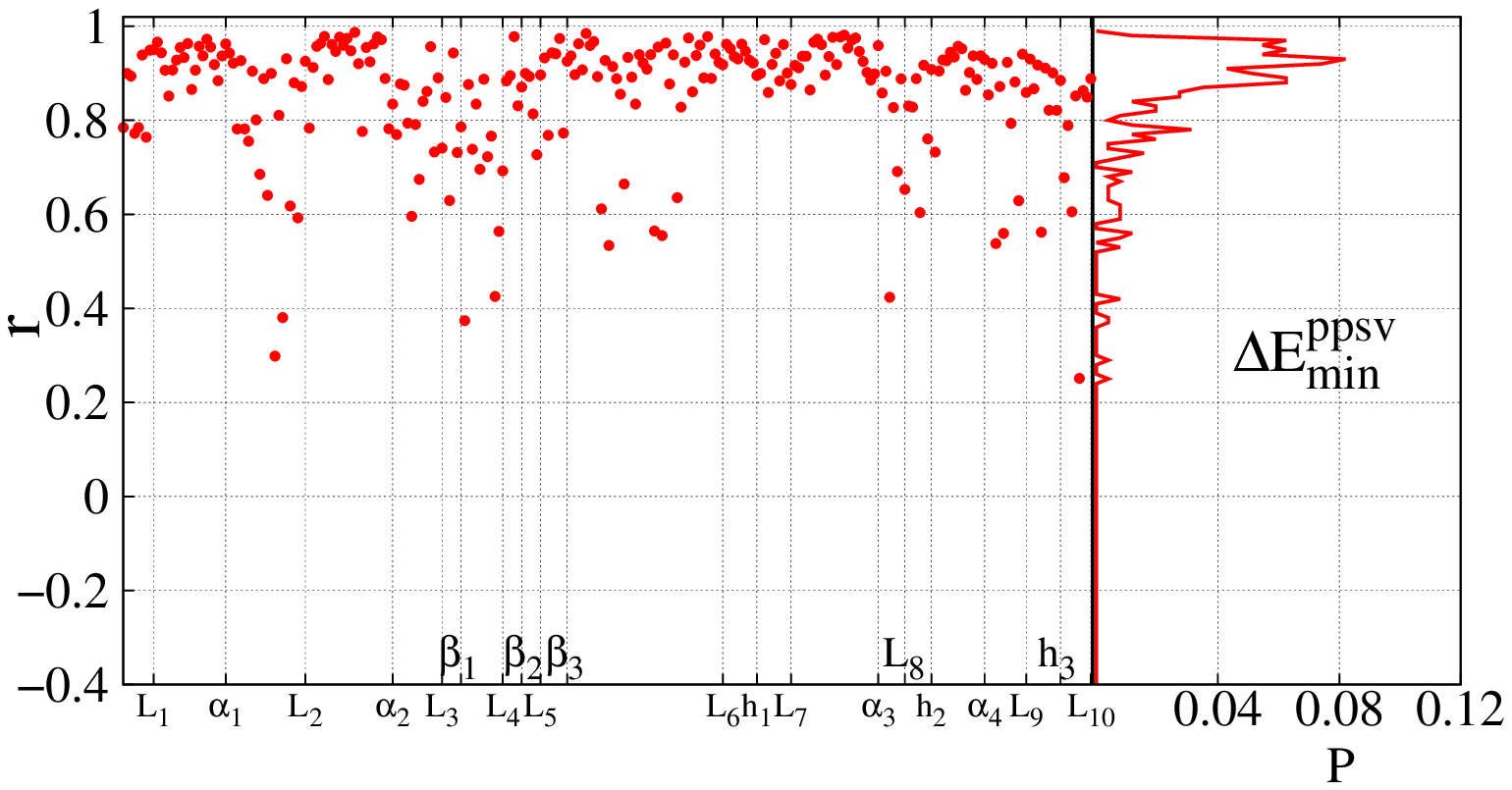}}
\subfloat[]{\includegraphics[width=1.8in]{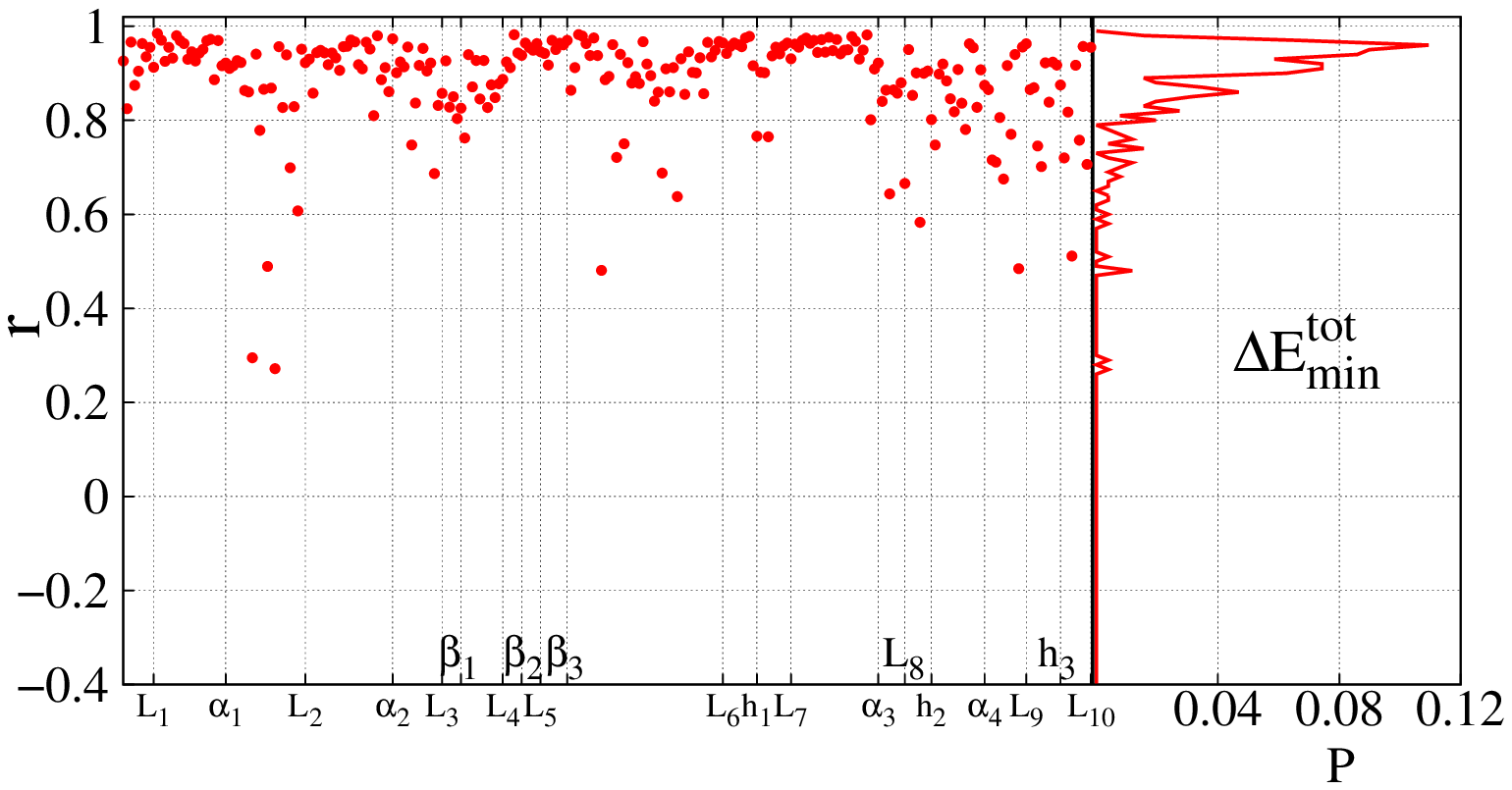}}\\
\caption{Linear correlations between $\Delta E^*_{min}$ (a $E^p_{min}$, b $E^{p\tn{-}sv}_{min}$, c $E^{ppsv}_{min}$, d $E^{tot}_{min}$) and $\Delta F$. Left panels are scatter plots of linear correlation coefficients ($r$) between $\Delta E^*_{min}$ and $\Delta F$ for macrostates pairs associated with each backbone dihedral, the corresponding secondary structures (see Fig. S1 for a graphic of secondary structures for HEWL) of backbone dihedrals are indicated on the horizontal axis. Right panels are the probability distributions of linear correlation coefficients observed in the 256 different ways of macrostates definition corresponding to 256 backbone dihedrals.}  
\label{fig:EminFE}
\end{figure}

\begin{figure}
\centering 
\subfloat[]{\includegraphics[width=1.8in]{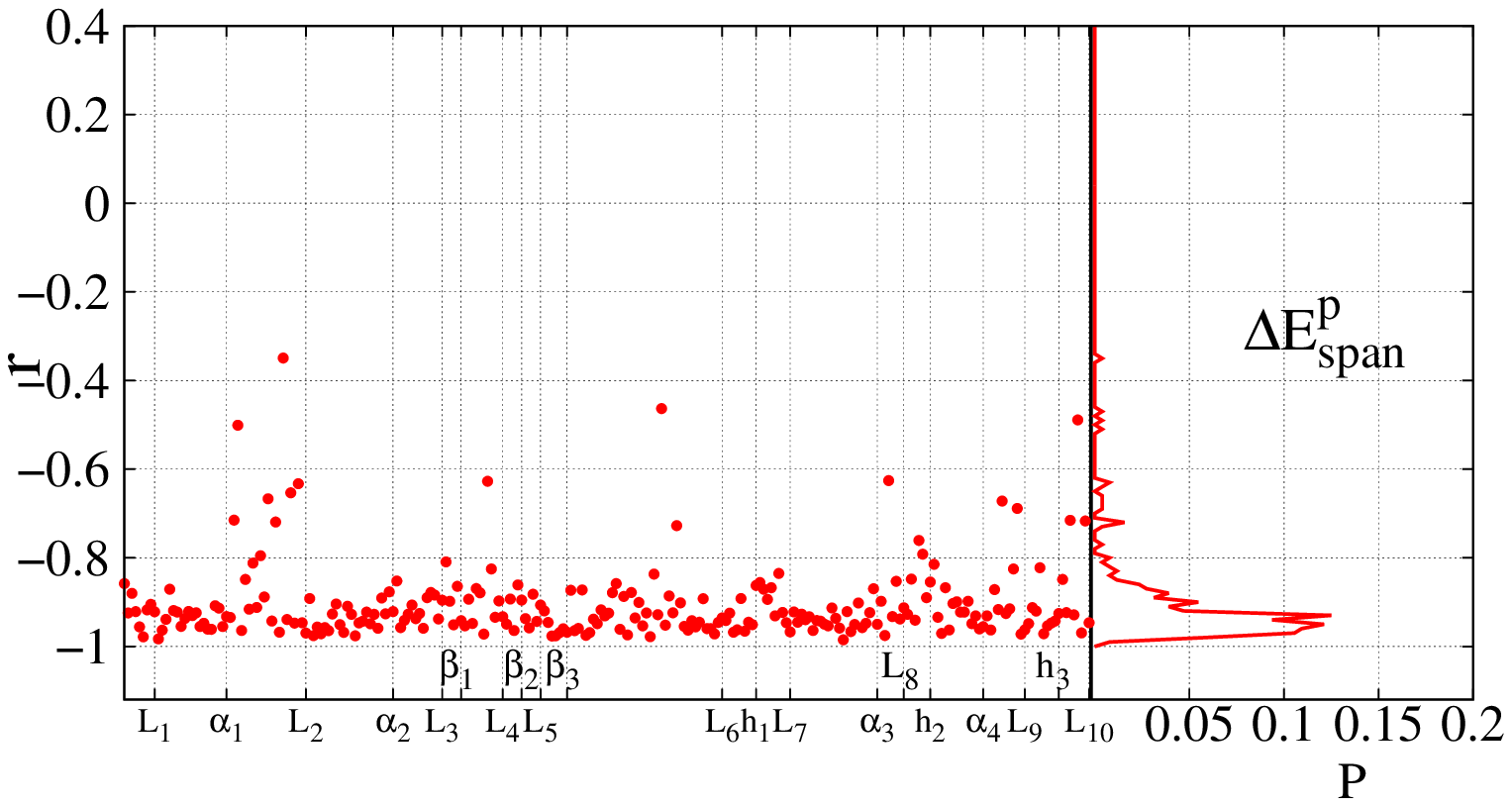}}
\subfloat[]{\includegraphics[width=1.8in]{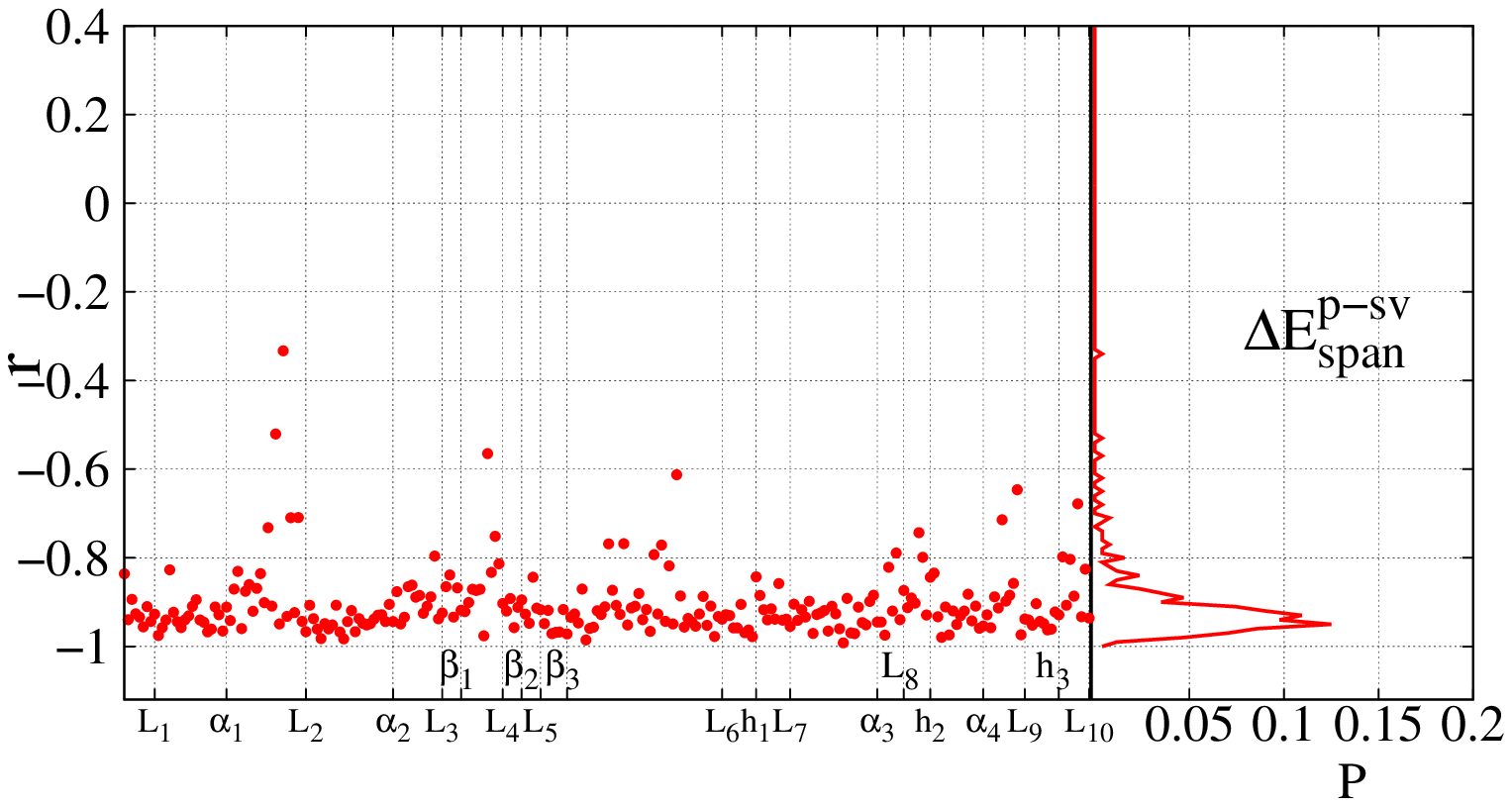}}\\
\subfloat[]{\includegraphics[width=1.8in]{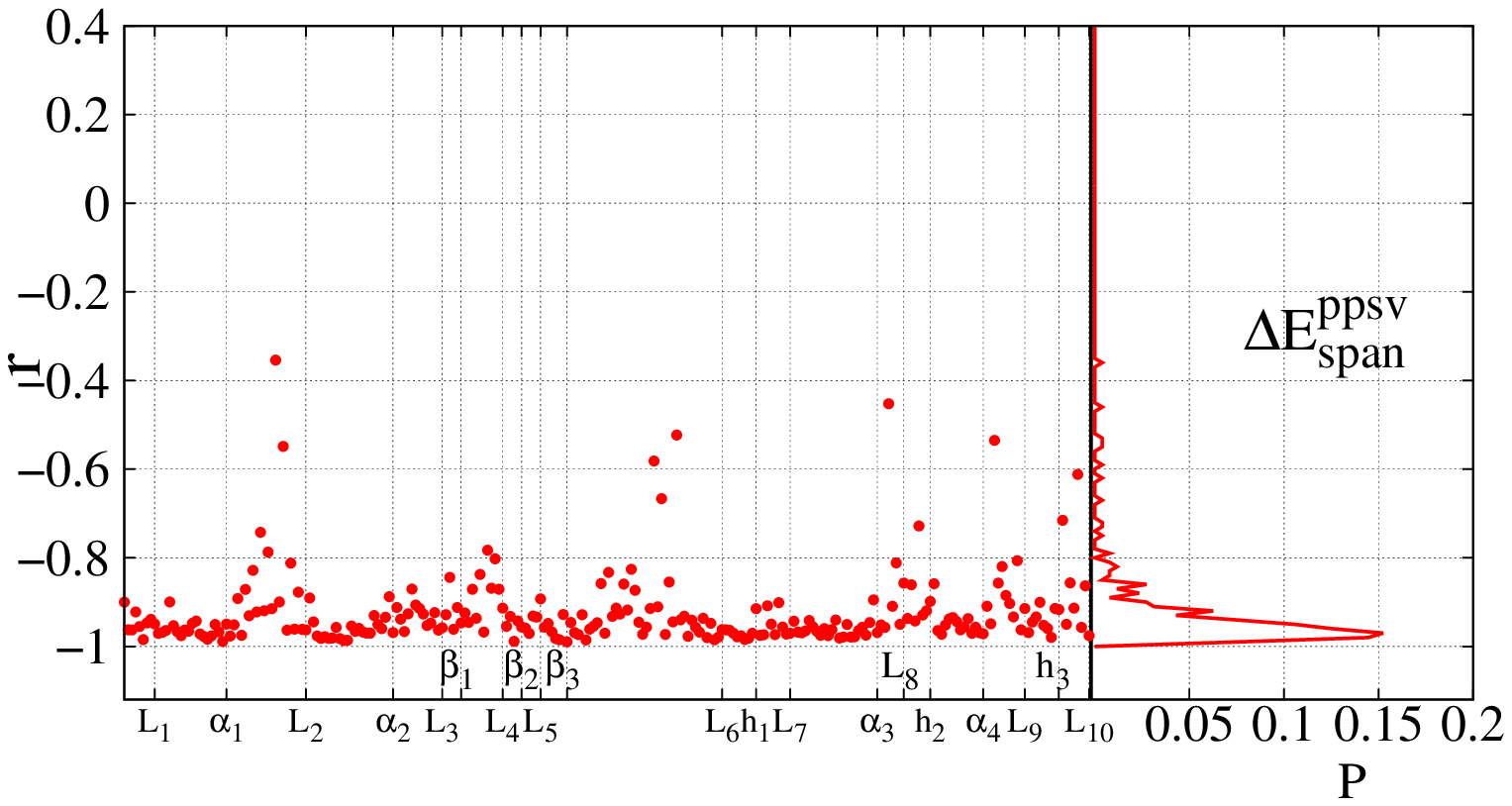}}
\subfloat[]{\includegraphics[width=1.8in]{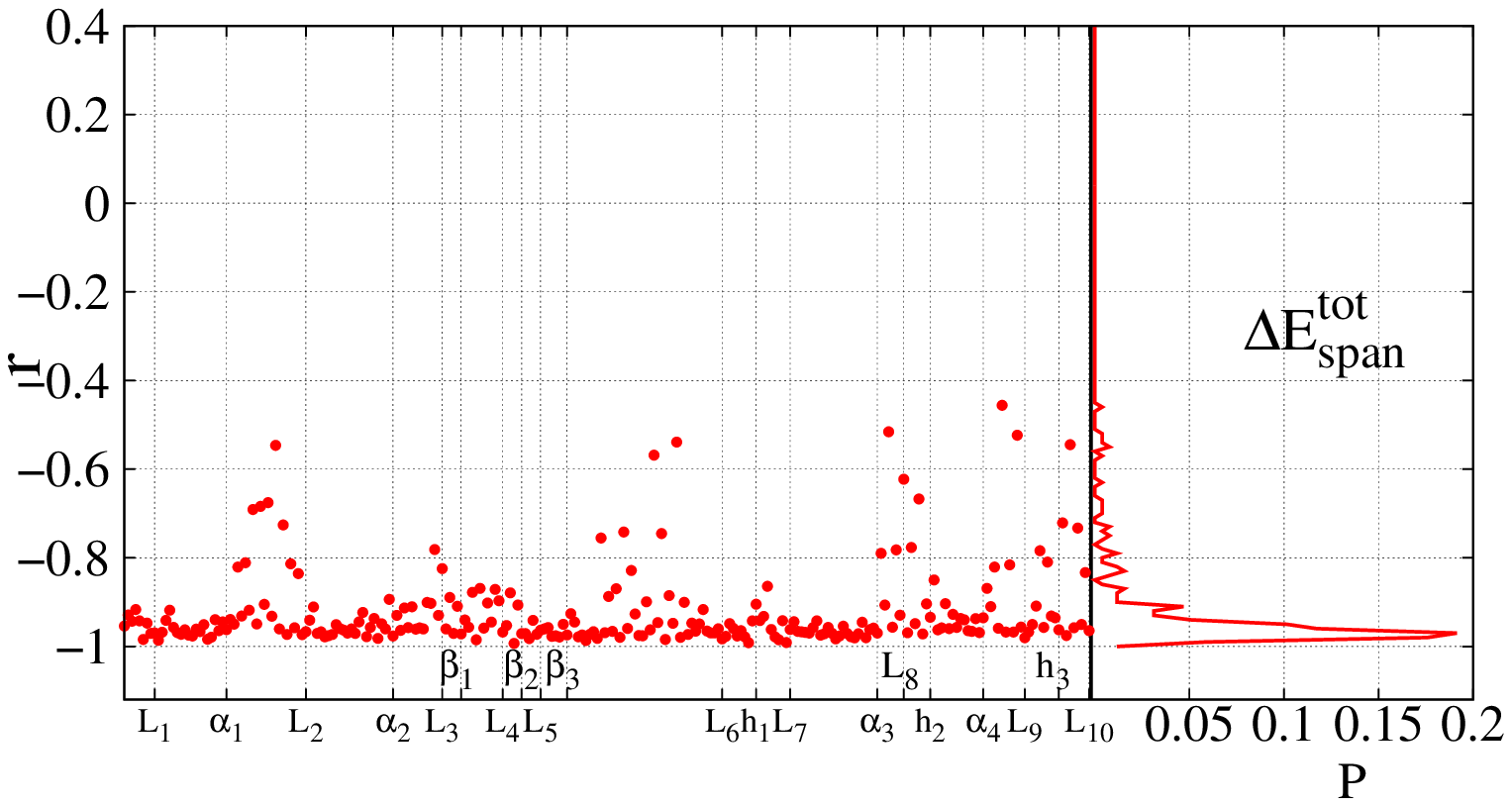}}\\
\caption{Linear correlations between $\Delta E^*_{span}$  (a $E^p_{span}$, b $E^{p\tn{-}sv}_{span}$, c $E^{ppsv}_{span}$ and d $E^{tot}_{span}$) and $\Delta F$. Left panels are the scatter plots of linear correlation coefficients ($r$) between $\Delta E^*_{span}$ and $\Delta F$ for macrostates associated with each backbone dihedral, the corresponding secondary structures of backbone dihedrals are indicated on the horizontal axis. Right panels are the probability distributions of linear correlation coefficients observed in the 256 different ways of macrostates definition corresponding to 256 backbone dihedrals.}
\label{fig:EspanFE}
\end{figure}

The apparent caveat of the MiPE approximation is neglect of entropic contributions. Based on the thought that MaPE terms, which are observed to correlate significantly with free energy, qualitatively reflect the entropic contributions in the high potential energy range, we figure that combination of MiPE and MaPE terms might accommodate both energetic and entropic contributions without increasing computational cost. Since change of MiPE terms positively correlate with change of free energy (Fig. \ref{fig:EminFE}), and change of MaPE terms negatively correlate with change of free energy (Fig. \ref{fig:EmaxFE}), $\Delta E^*_{span} = \Delta (E^*_{max} - E^*_{min})$ are potentially better free energy proxies than corresponding change of both MiPE and MaPE terms. We calculated $\Delta E^*_{span}$ and correlations of these terms with $\Delta F$, and the results are presented in Fig. \ref{fig:EspanFE}. Firstly, all $\Delta E^*_{span}$ terms indeed strongly correlate with $\Delta F$, with $\Delta E^{tot}_{span}$ and $\Delta E^{ppsv}_{span}$ exhibit stronger correlations than $\Delta E^p_{span}$ and $\Delta E^{p\tn{-}sv}_{span}$ (Fig.\ref{fig:CPDmmsa})c. 
More importantly, as suspected, $\Delta E^*_{span}$ correlated with $\Delta F$ consistently and significantly better than both $\Delta E^*_{min}$ and $\Delta E^*_{max}$, regardless of the specific dihedrals utilized to define macrostates and potential energy components utilized (Fig. \ref{fig:CPDpppt}). Similar to analysis of $\Delta E^*_{min}$, macrostates defined by different backbone dihedrals were pooled together (with 13104640 macrostate pairs in total) to perform an overall linear fit, the results were shown in the supporting Fig. S\ref{fig:EspanFE}.  

\section*{\sffamily \Large Correlations between change of average potential energy terms and change of free energy for HEWL}
\begin{figure}
\centering 
\subfloat[]{\includegraphics[width=1.5in]{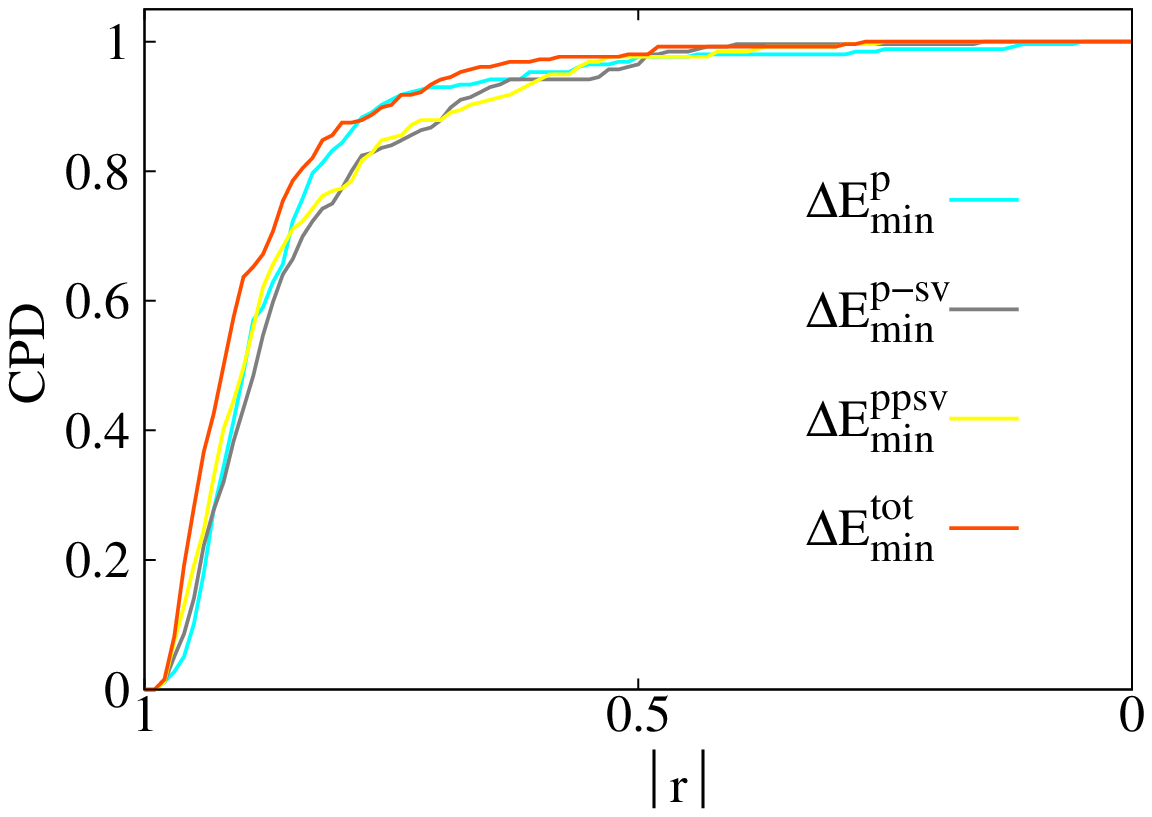}}
\subfloat[]{\includegraphics[width=1.5in]{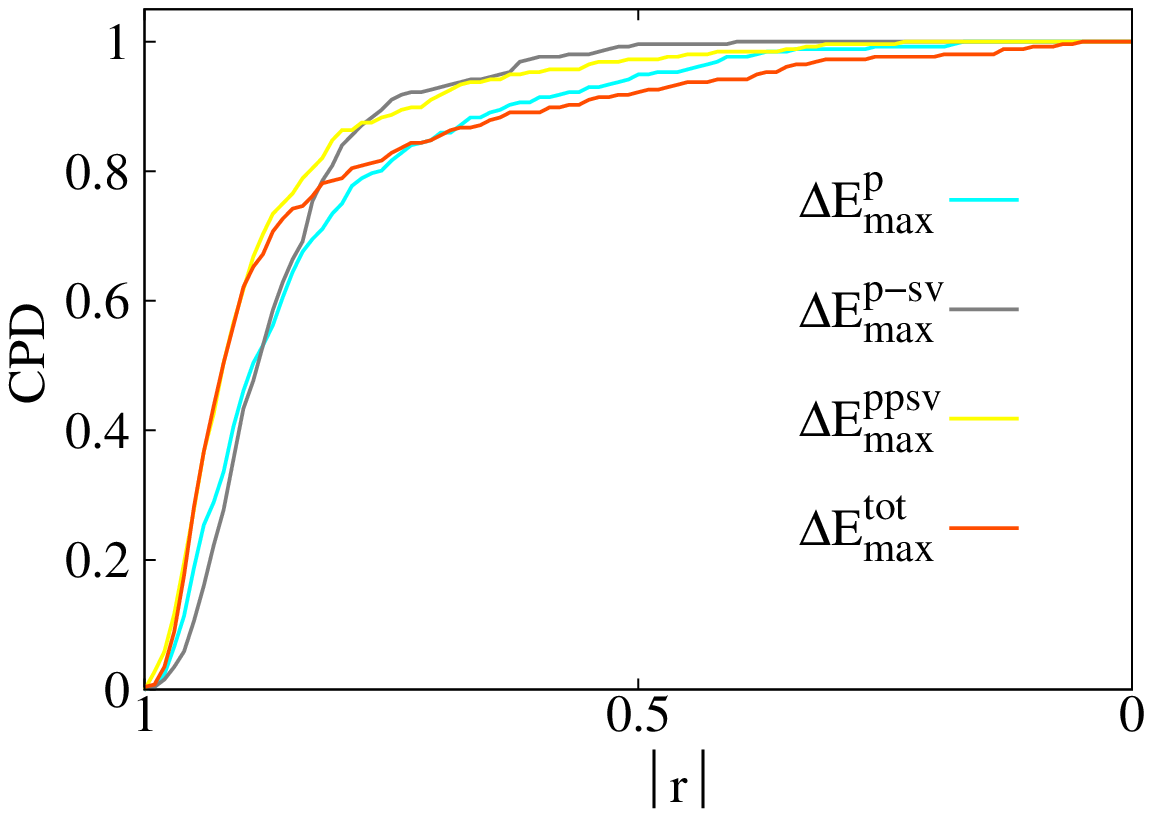}}\\
\subfloat[]{\includegraphics[width=1.5in]{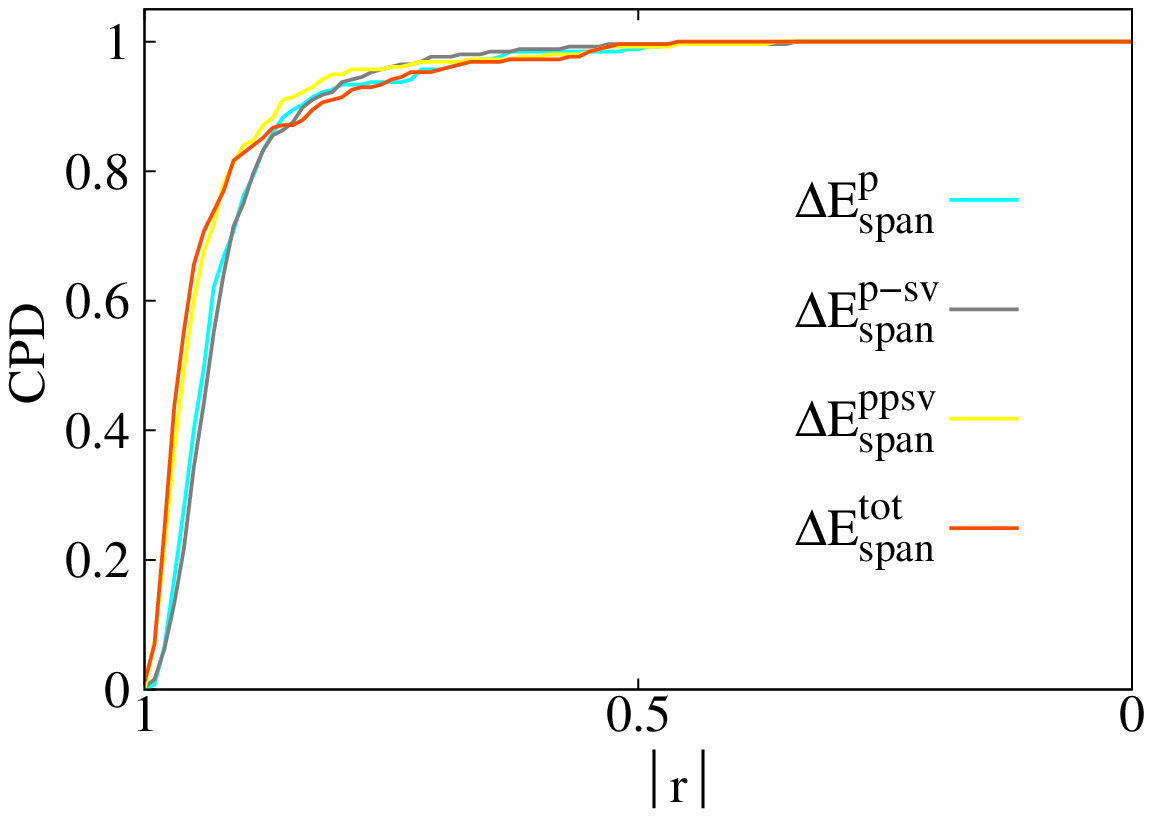}}
\subfloat[]{\includegraphics[width=1.5in]{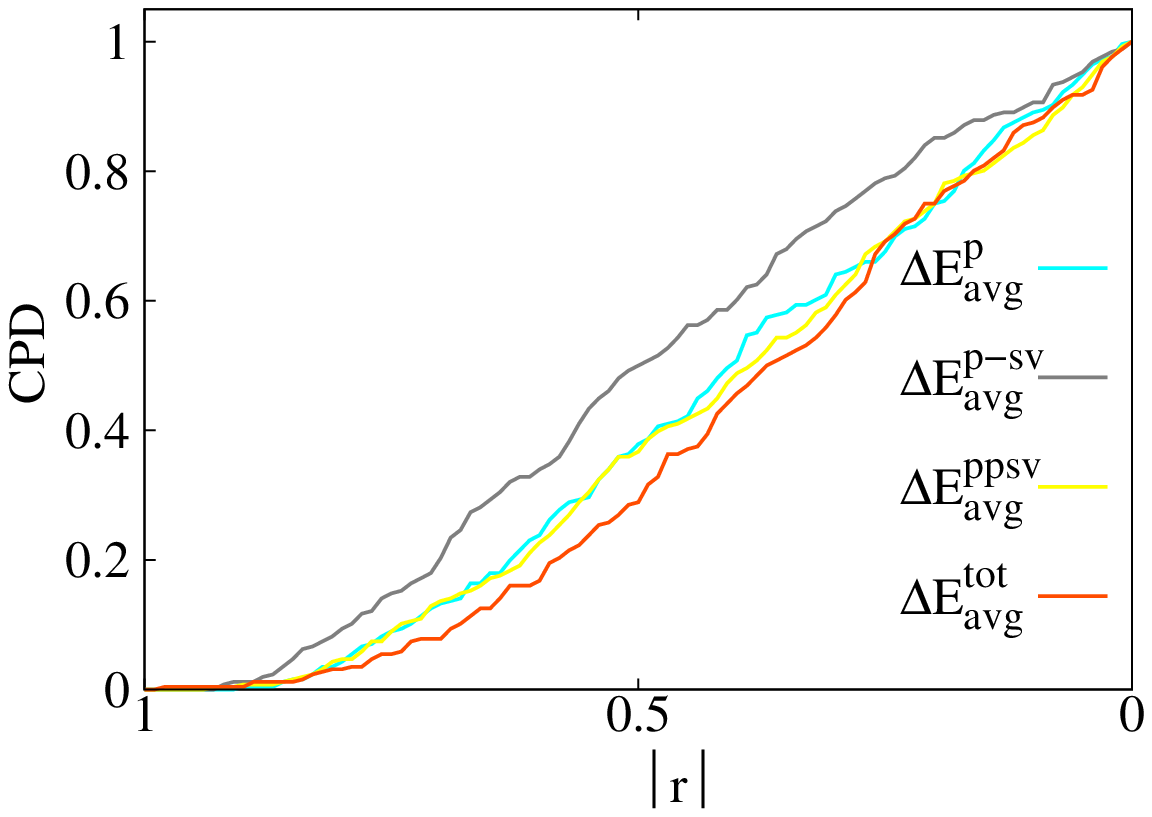}}
\caption{The cumulated probability density (CPD) of linear correlation coefficients between a $\Delta E^*_{min}$, b $\Delta E^*_{max}$, c $\Delta E^*_{span}$, d $\Delta E^*_{avg}$ and $\Delta F$  as a function of decreasing $|r|$ from 1 to 0.}
\label{fig:CPDmmsa}
\end{figure}

\begin{figure}
\centering 
\subfloat[]{\includegraphics[width=1.8in]{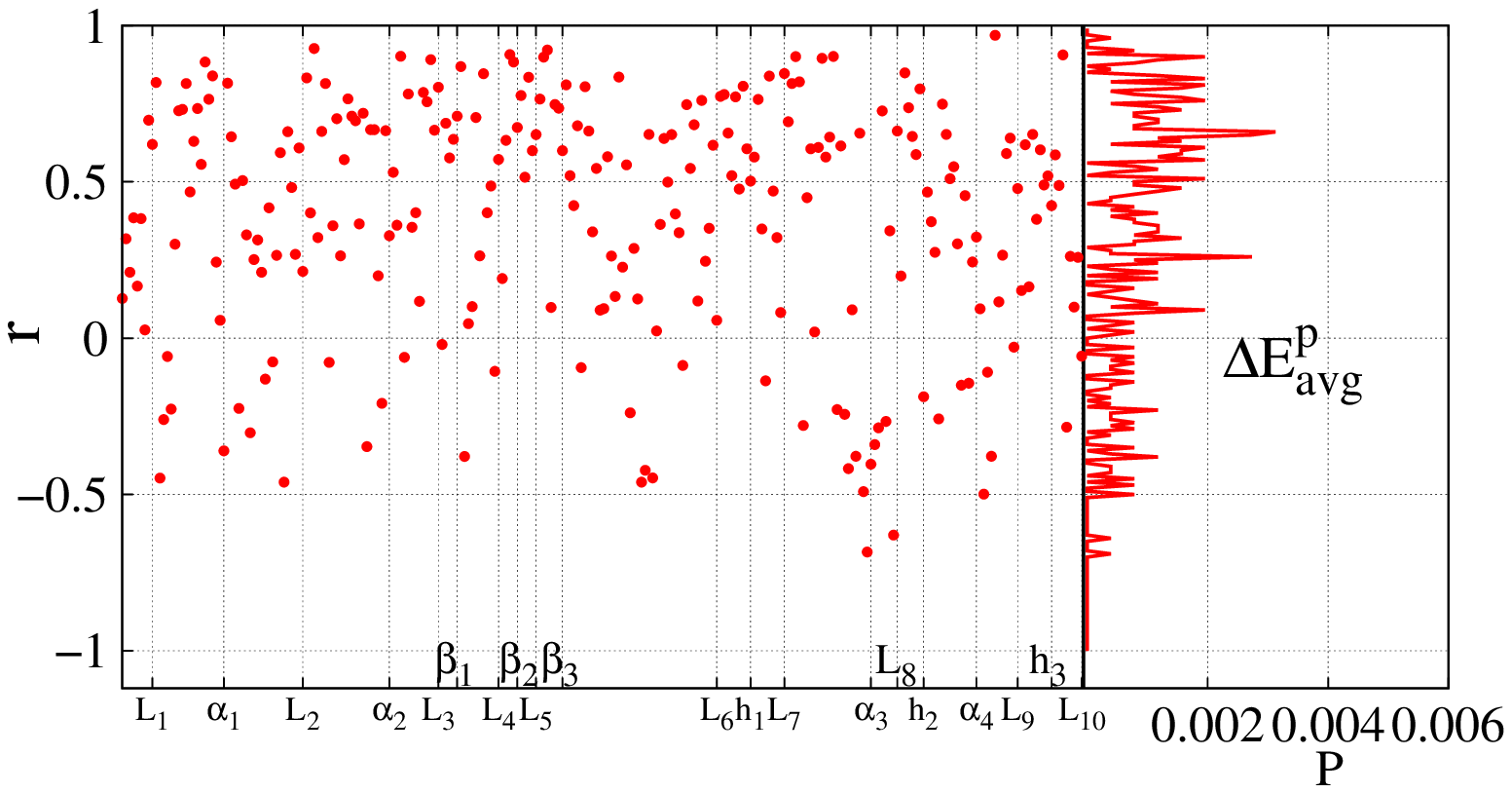}}
\subfloat[]{\includegraphics[width=1.8in]{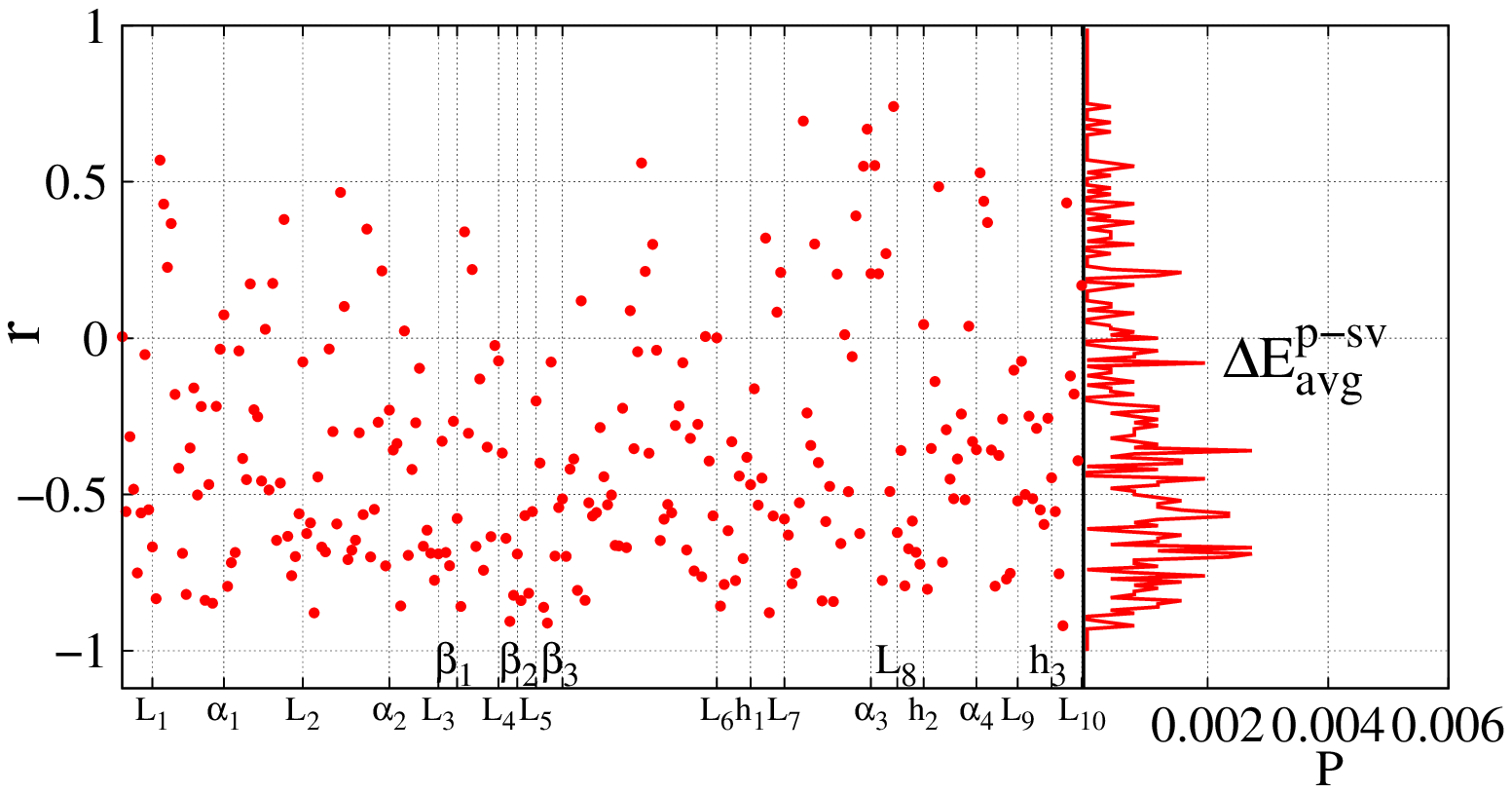}}\\
\subfloat[]{\includegraphics[width=1.8in]{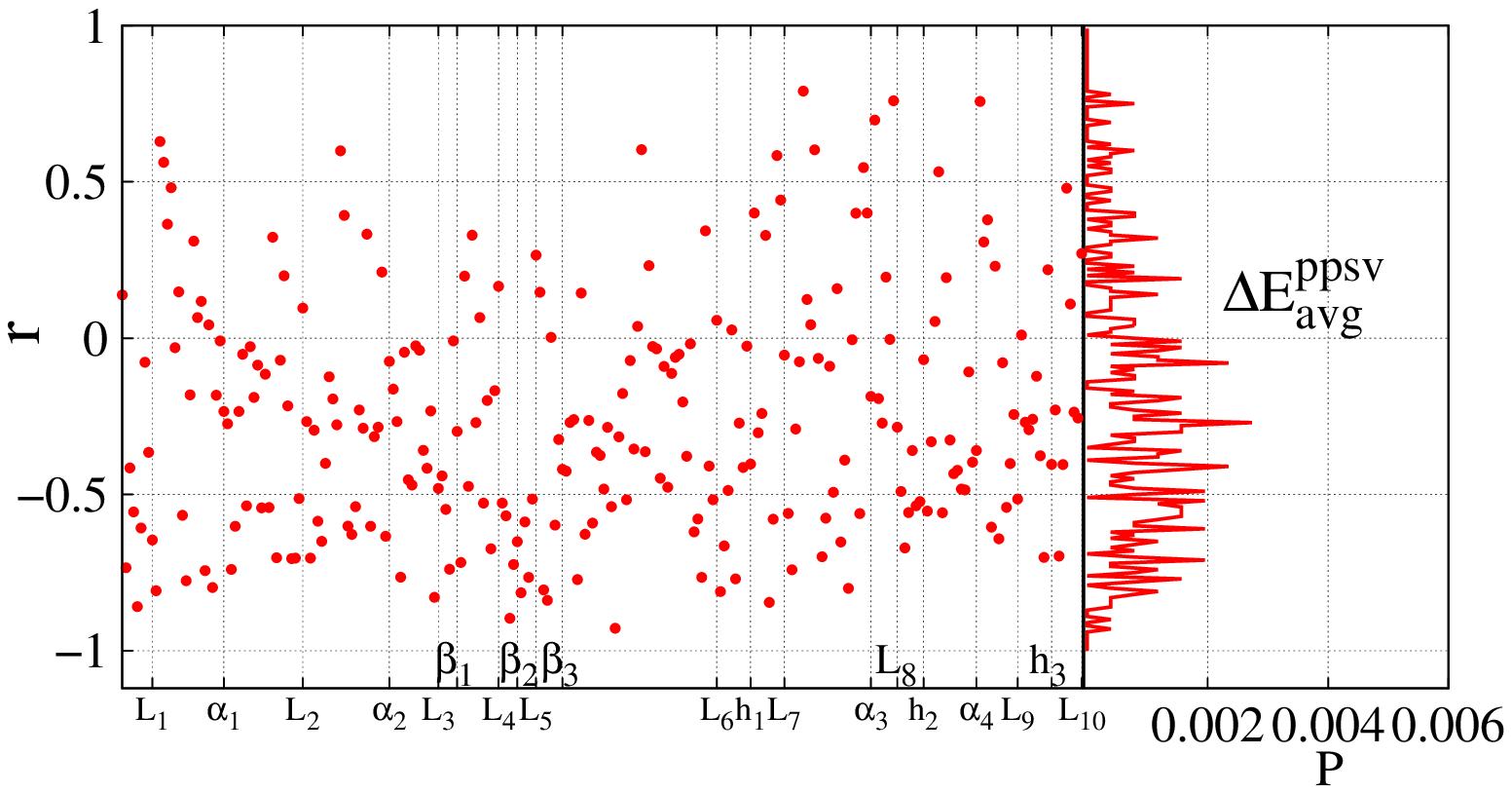}}
\subfloat[]{\includegraphics[width=1.8in]{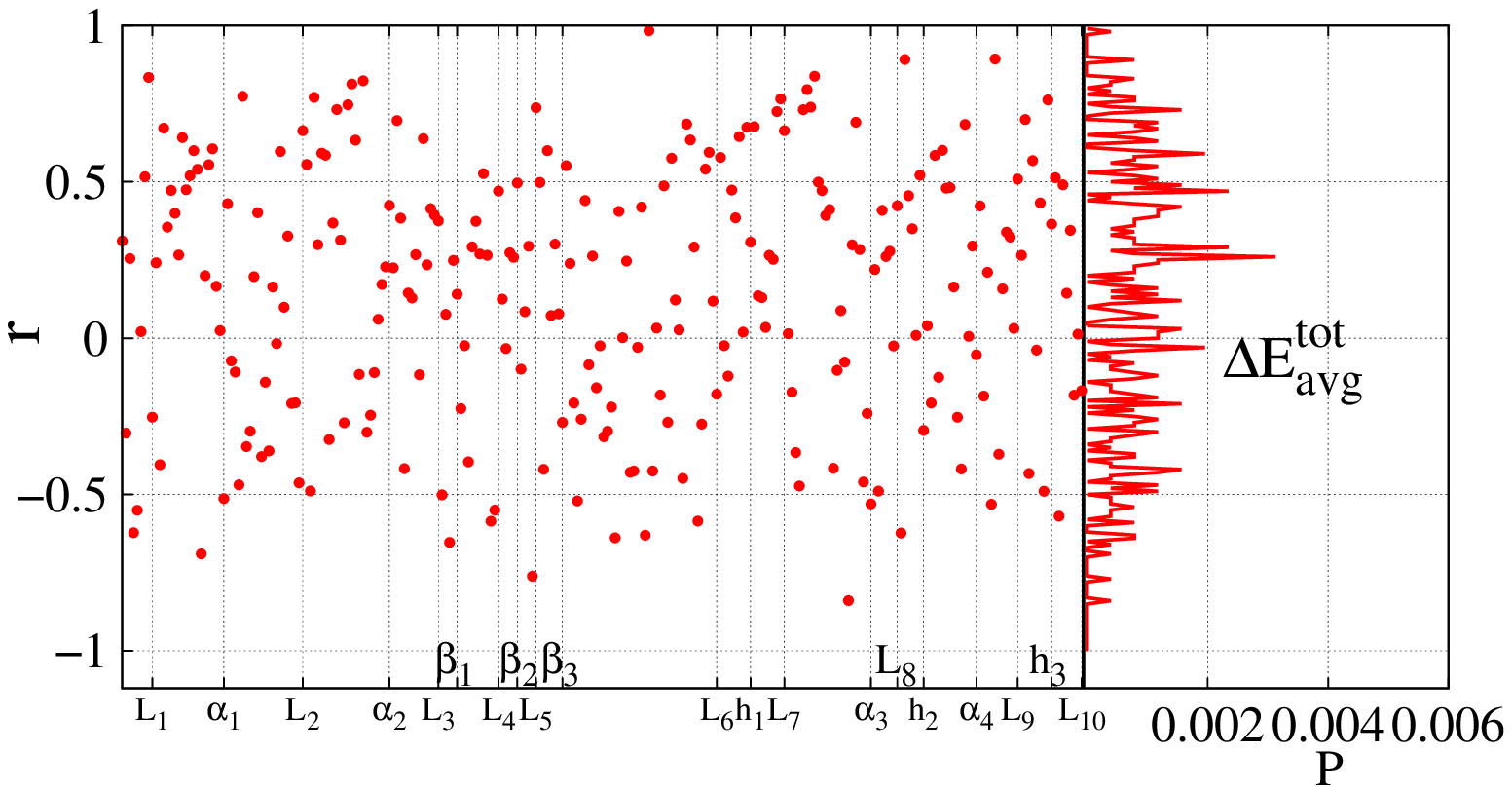}}
\caption{Linear correlations between $\Delta E^*_{avg}$  (a $E^p_{avg}$, b $E^{p\tn{-}sv}_{avg}$, c $E^{ppsv}_{avg}$ and d $E^{tot}_{avg}$) and $\Delta F$. Left panels are scatter plots of linear correlation coefficients ($r$) between $\Delta E^*_{avg}$ and $\Delta F$ for macrostates associated with each backbone dihedral, the corresponding secondary structures of backbone dihedrals are indicated on the horizontal axis. Right panels are the probability distributions of linear correlation coefficients observed in the 256 different ways of macrostates definition corresponding to 256 backbone dihedrals.}
\label{fig:EavgFE}
\end{figure}

It is widely believed that macrostates with lower average potential energy are likely to have lower free energy. Change of ensemble averaged potential energy term is essential when the basic formula for change of free energy
\begin{equation} 
\Delta F = \Delta E - T\Delta S
\label{eq:DFbasic}
\end{equation}
is utilized as in the case of MM/P(G)BSA. To test this intuitive belief, we calculated $\Delta E^*_{avg}$ (the subscript $_{avg}$ represents ensemble average based on MD trajectory set.) and correlations of these terms with change of free energy were presented in Fig. \ref{fig:EavgFE}. It is clear that change of all average potential energy terms essentially have no significant correlations with change of free energy. This is mathematically a direct result of negative correlations between $\Delta E^*_{max}$ and $\Delta F$ and positive correlations between $\Delta E^*_{min}$ and $\Delta F$. The process of averaging is essentially adding up two sides of a distribution and the opposing effects of these two sides cancel each other. The finding suggests that direct utilization of equation \ref{eq:DFbasic} requires accurate estimation of $\Delta S$, which is acknowledged as an extremely challenging task\cite{}. It is suggested that due to almost ubiquitous existence of entropy enthalpy compensation phenomenon\cite{Chodera2013}, straight forward calculation of change of free energy with equation \ref{eq:DFbasic} is likely to be unreliable. An alternative perspective for explaining this observation is that a given average potential energy may be obtained either from a sharply peaked narrow distribution or a flat wide distribution or anything in between. The observed strong correlations between $\Delta E^*_{span}$ and $\Delta F$ is consistent with weak correlations between $\Delta E^*_{avg}$ and $\Delta F$. For $\Delta E^*_{min}$, $\Delta E^*_{max}$ and $\Delta E^*_{span}$, the ``$tot$'' and ``$ppsv$'' terms exhibit stronger correlations with $\Delta F$ than ``$p$'' and ``$p\tn{-}sv$'' terms (Fig. \ref{fig:CPDpppt}a, b and c; Fig. S2). This makes intuitive sense since ``$tot$'' and ``$ppsv$'' terms includes more potential energy components, all of which contribute to free energy. The opposite is observed for $\Delta E^*_{avg}$ (Fig. \ref{fig:CPDmmsa}d). However, since all $\Delta E^*_{avg}$ terms correlate weakly with $\Delta F$, the difference among them are likely not meaningful both theoretically and practically.

\section*{\sffamily \Large DISCUSSION}
\begin{figure}
\centering 
\subfloat[]{\includegraphics[width=1.8in]{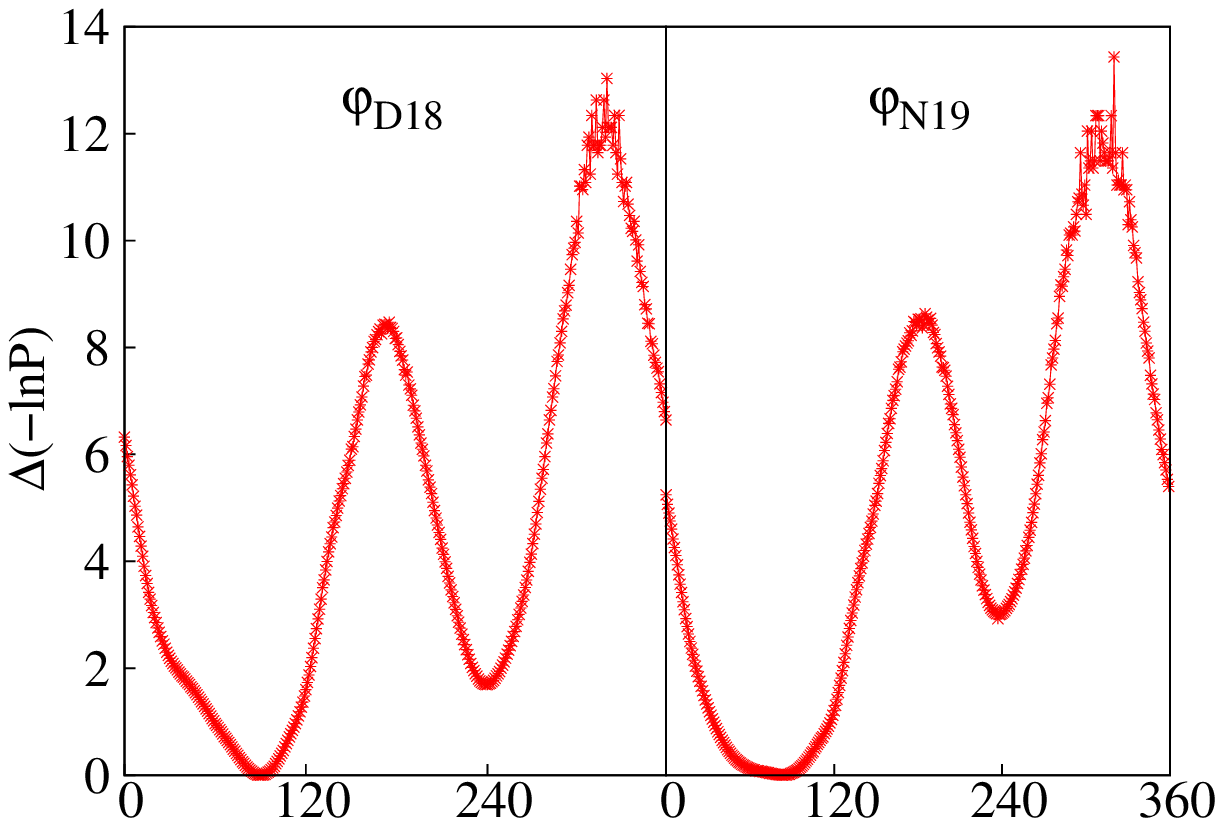}}
\subfloat[]{\includegraphics[width=1.8in]{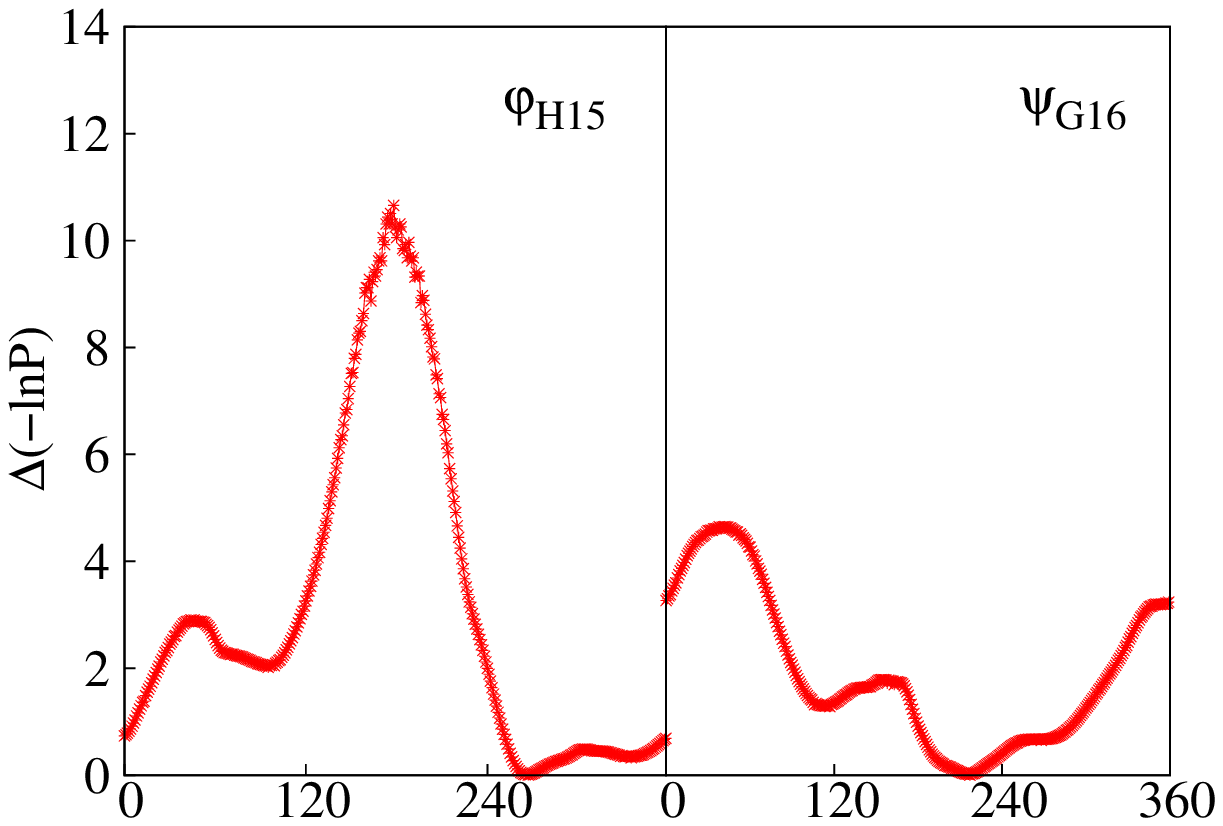}}\\
\subfloat[]{\includegraphics[width=1.8in]{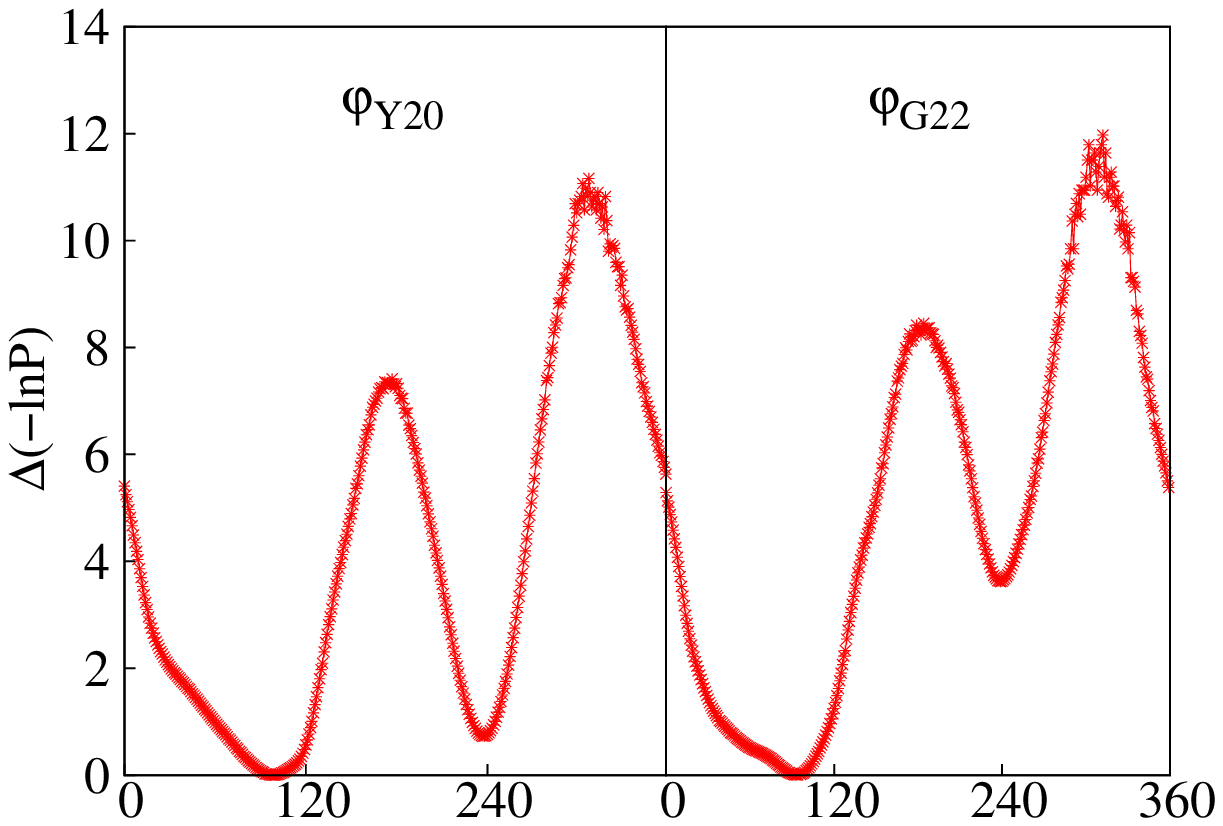}}
\subfloat[]{\includegraphics[width=1.8in]{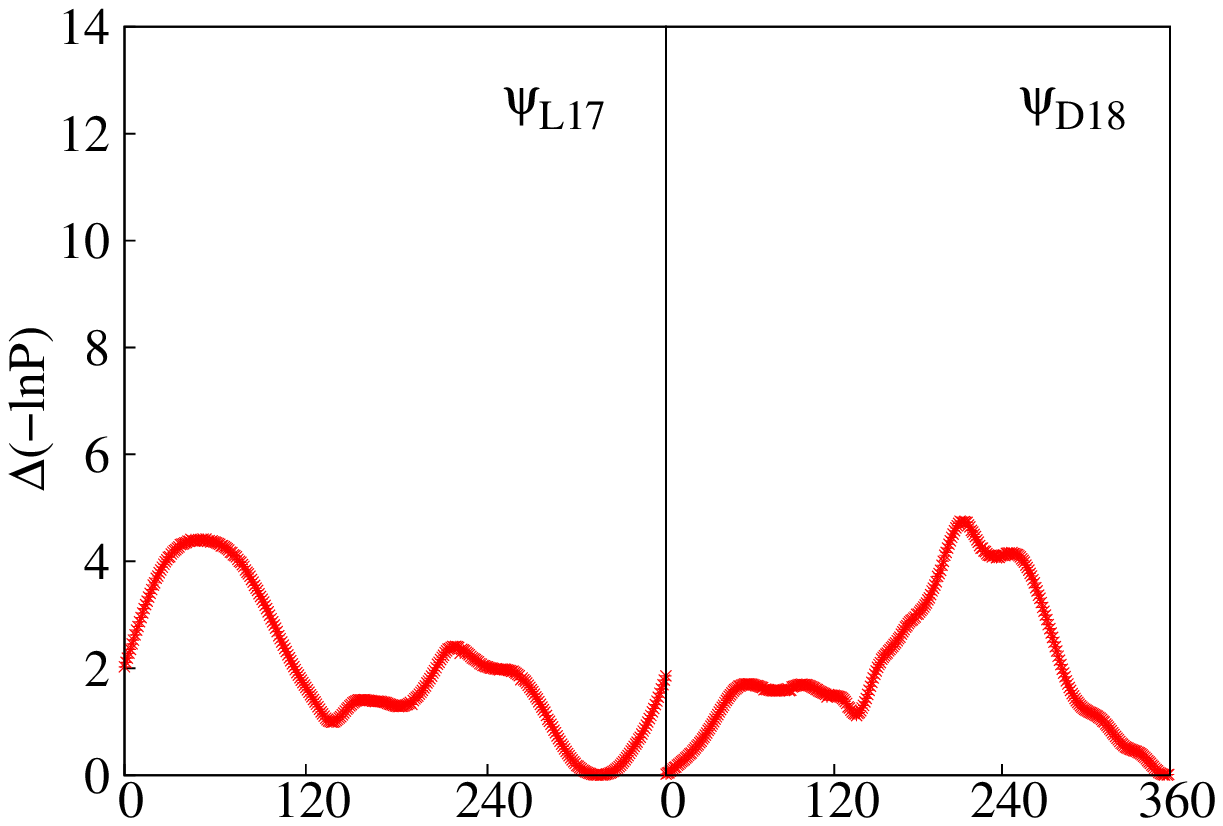}}\\
\subfloat[]{\includegraphics[width=1.8in]{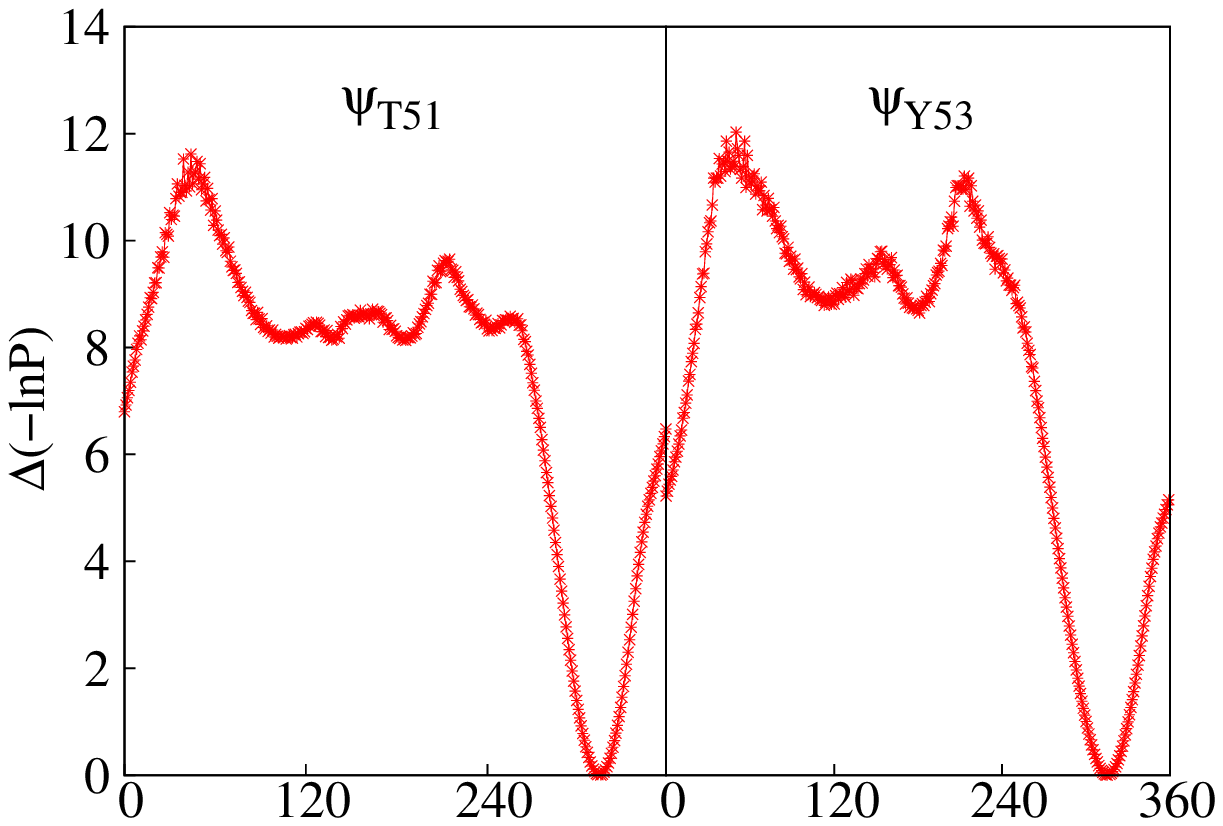}}
\subfloat[]{\includegraphics[width=1.8in]{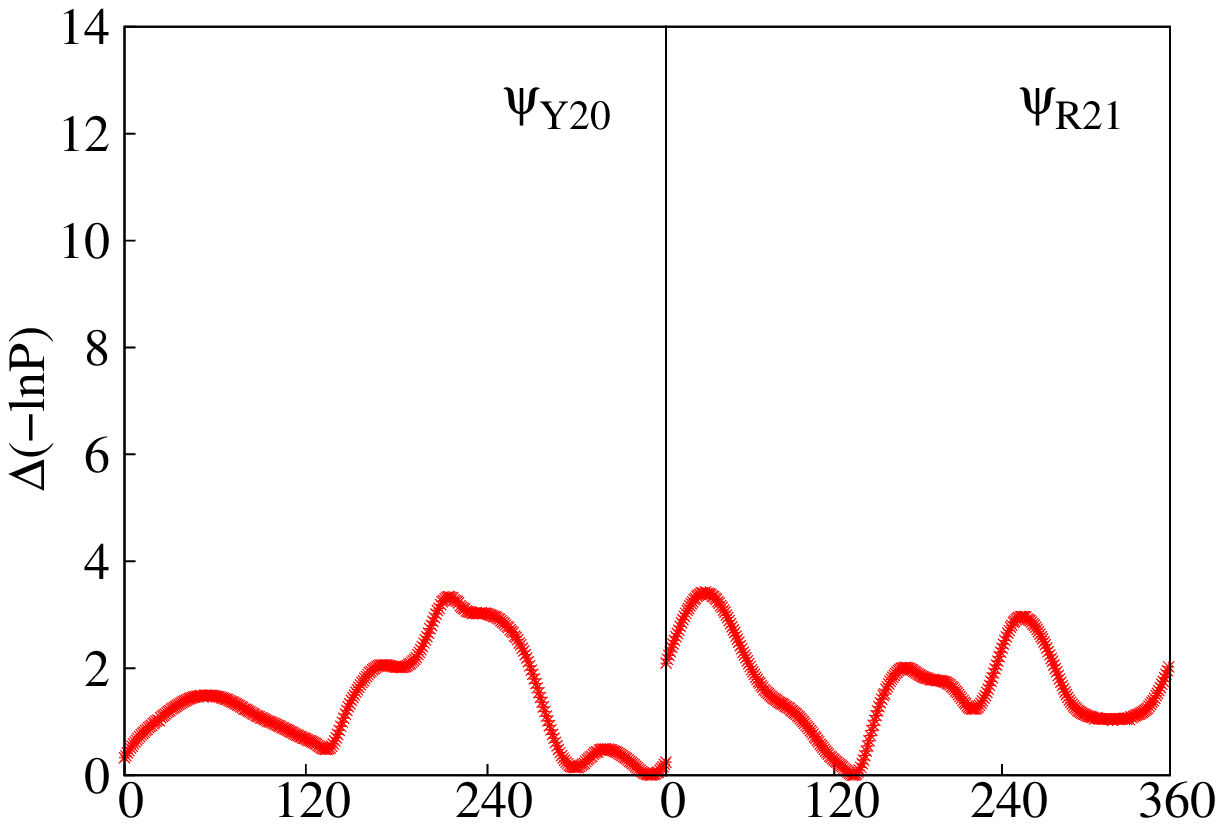}}\\
\subfloat[]{\includegraphics[width=1.8in]{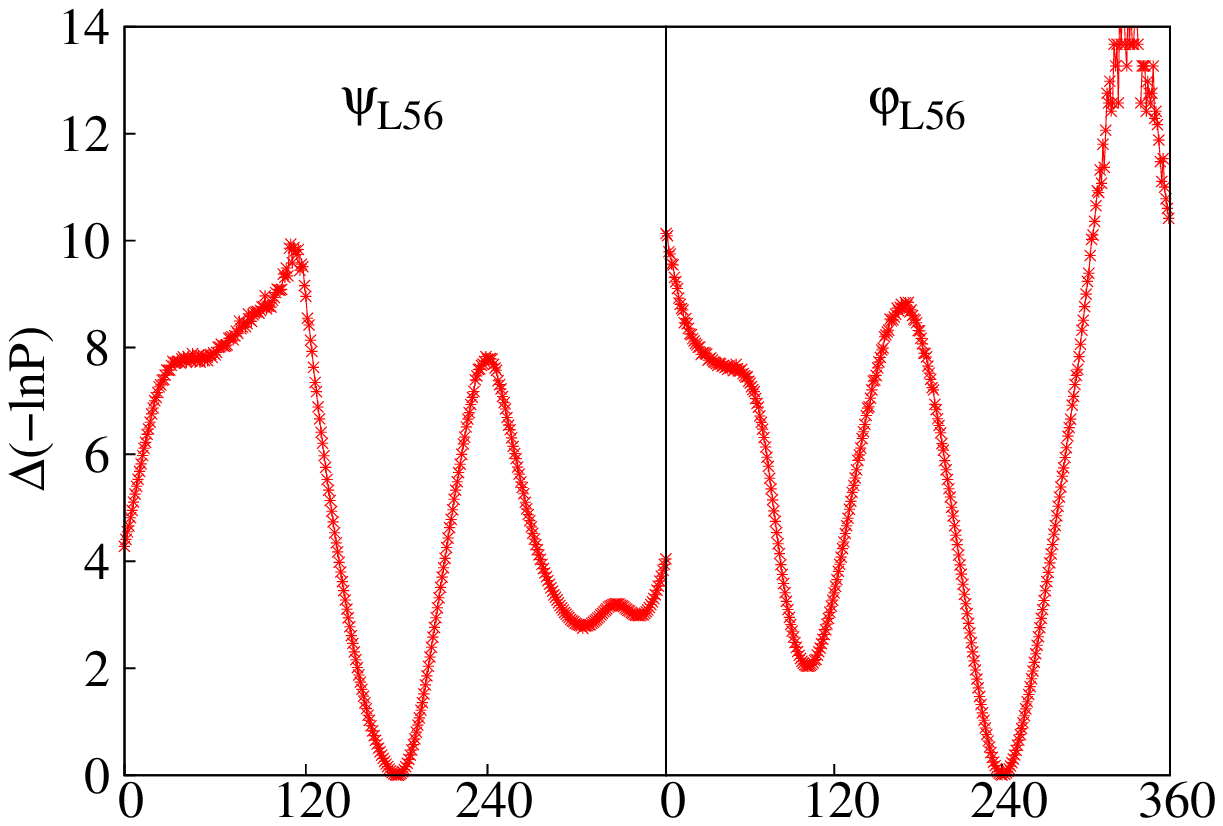}}
\subfloat[]{\includegraphics[width=1.8in]{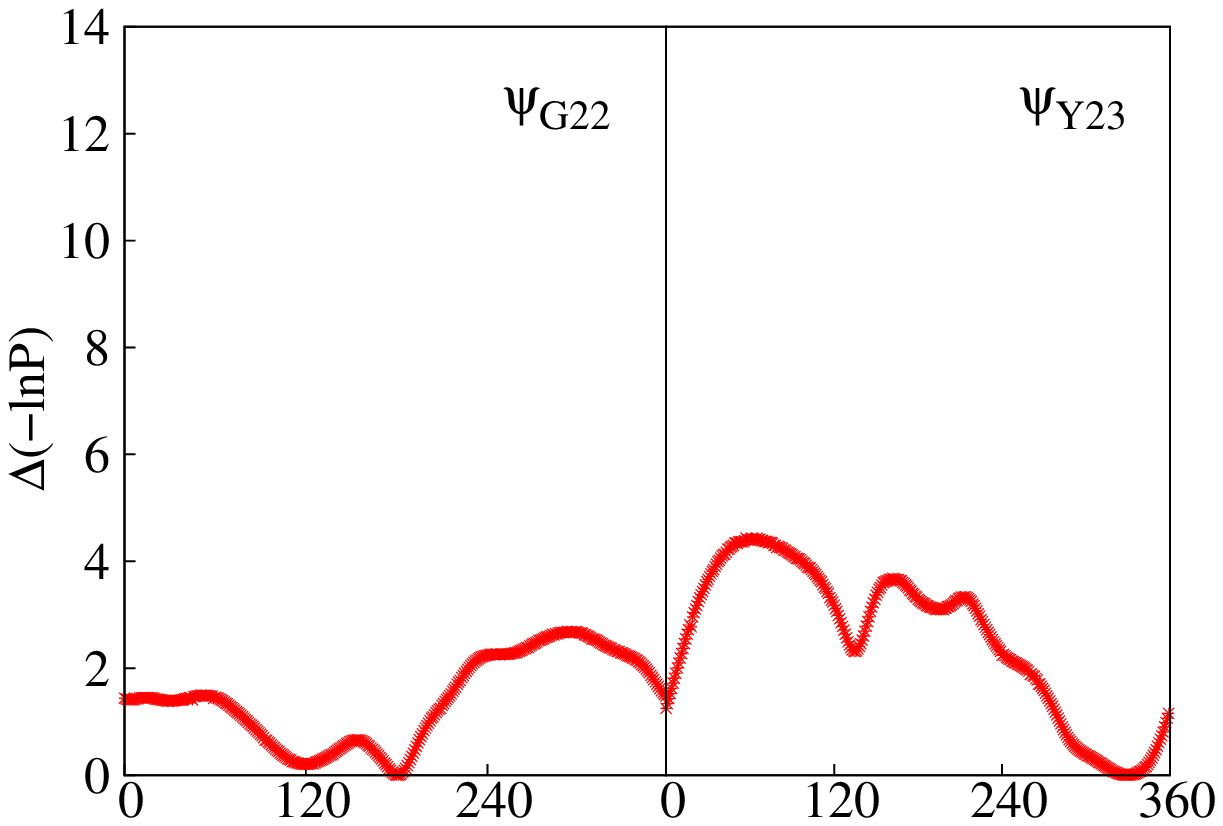}}
\caption{Adjusted negative natural logarithm probability ($\Delta (-lnP)$) of the representative SCLDs (a, c, e, g) and WCLDs (b, d, f, h), the identity of each plotted backbone dihedral is labeled (e.g. $\phi _{19R}$). $\Delta (-lnP)$ is obtained as follows, each dihedral is divided into 360 $1^\circ$ bins and their negative natural log probability calculated and adjusted by subtracting the smallest value $-lnP_{min}$. Therefore, $\Delta (-lnP)$ is relative free energy projected onto each backbone dihedrals based on $1^\circ$ bins. }
\label{fig:SWCLD}
\end{figure}

For all potential energy terms (except $E^*_{avg}$) analyzed, a consistent pattern is that for macrostates defined by projection onto some backbone dihedrals in loop regions, significantly weaker correlations are observed between $\Delta E^*_{min}$, $\Delta E^*_{max}$, $\Delta E^*_{span}$ and $\Delta F$ than that of macrostates defined by all other backbone dihedrals. To identify cause of such difference in correlations between change of potential energy terms and change of free energy, we divided backbone dihedrals in loop region into three types. The first type of dihedrals are those utilized to define macrostates exhibiting strong linear correlations between $\Delta E^*_{min, max, span}$ and $\Delta F$ (with all three linear correlation coefficients greater than 0.9, hereafter addressed as strongly correlating loop dihedrals(SCLD)), and the second type are those utilized to define macrostates exhibiting weak corresponding linear correlations (with all three linear correlation coefficients smaller than 0.6, hereafter addressed as weakly correlating loop dihedrals(WCLD)), the remaining are classified as moderately correlating loop dihedrals (MCLD). Adjusted negative natural logarithm probability (see  Fig. \ref{fig:SWCLD} for details), which is effectively free energy with the lowest point as the reference, for representative dihedrals of both SCLDs and WCLDs were shown in Fig. \ref{fig:SWCLD}. 

It is immediately clear that high barriers dividing SCLD into local wells (Fig.\ref{fig:SWCLD}a, c, e, g). For most WCLDs, there are no significant free energy barriers at all (Fig. \ref{fig:SWCLD}d, f, h and $\psi _{G16}$ in b). While $\phi _{H15}$ in Fig. \ref{fig:SWCLD}b) has one high free energy barrier, which does not hinder rapid diffusion along the dihedral due to the fact that dihedrals are cyclic. Therefore, WCLDs are essentially (nearly) freely rotating in the whole range, with time scales of diffusion along these dihedrals being on nano-seconds or shorter as revealed by examination of original MD trajectories. While transitions between local wells for SCLDs occur on sub-micron seconds or longer than time scales. Physically, each macrostate defined by projection onto the slowest degrees of freedom (DOFs) (e.g. dihedrals with long average transition times between different torsional states) likely comprise a continuous region or a few discrete regions in configurational space. In contrast, each macrostate defined by projection onto very fast DOFs generally comprising a great number of discrete and distal fragments in configurational space, and calculating change of free energy between such two ``macrostates'' is both difficult and meaningless in reality. Therefore, weak correlations observed between change of potential energy terms $\Delta E^*_{min, max, span}$ should not be a great concern for physically well-defined macrostates (continuous regions in configurational space). This fact, on the other hand, remind us to be careful in defining macrostates when exploring free energy landscape of complex molecular systems.

\begin{figure}
\centering 
\subfloat[]{\includegraphics[width=1.5in]{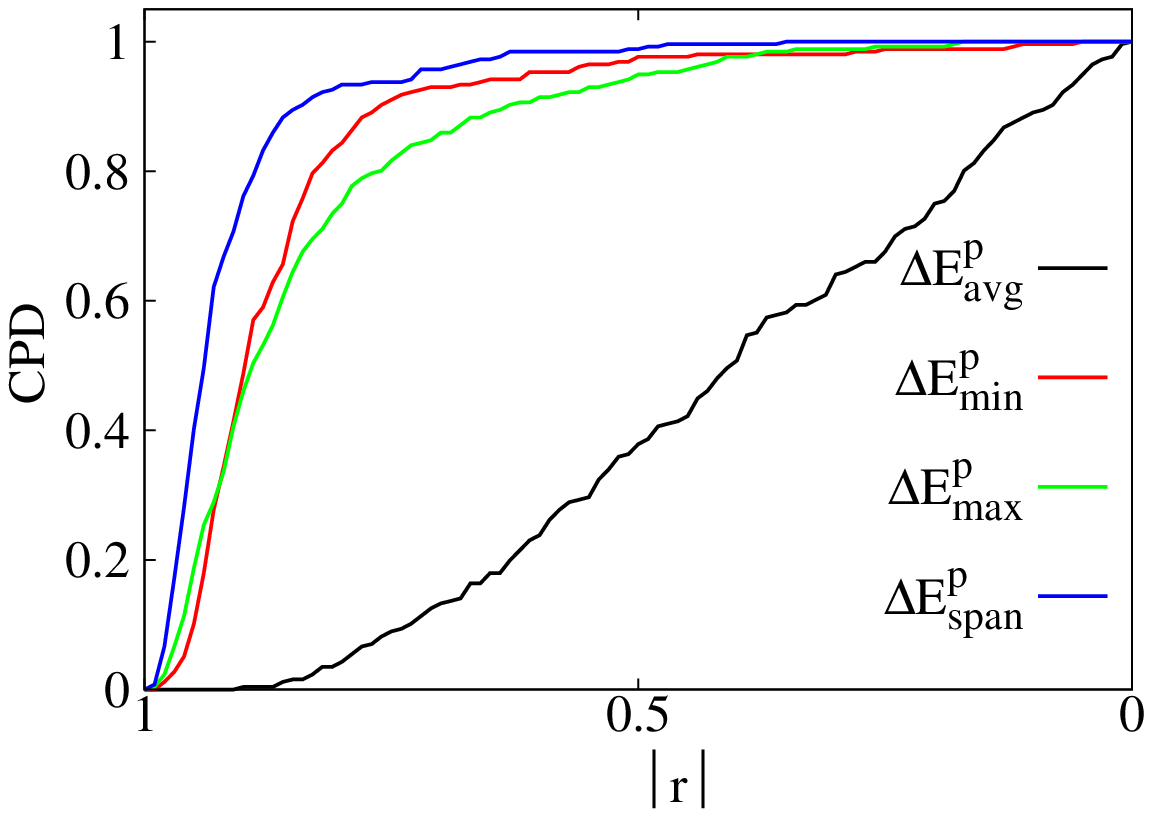}}
\subfloat[]{\includegraphics[width=1.5in]{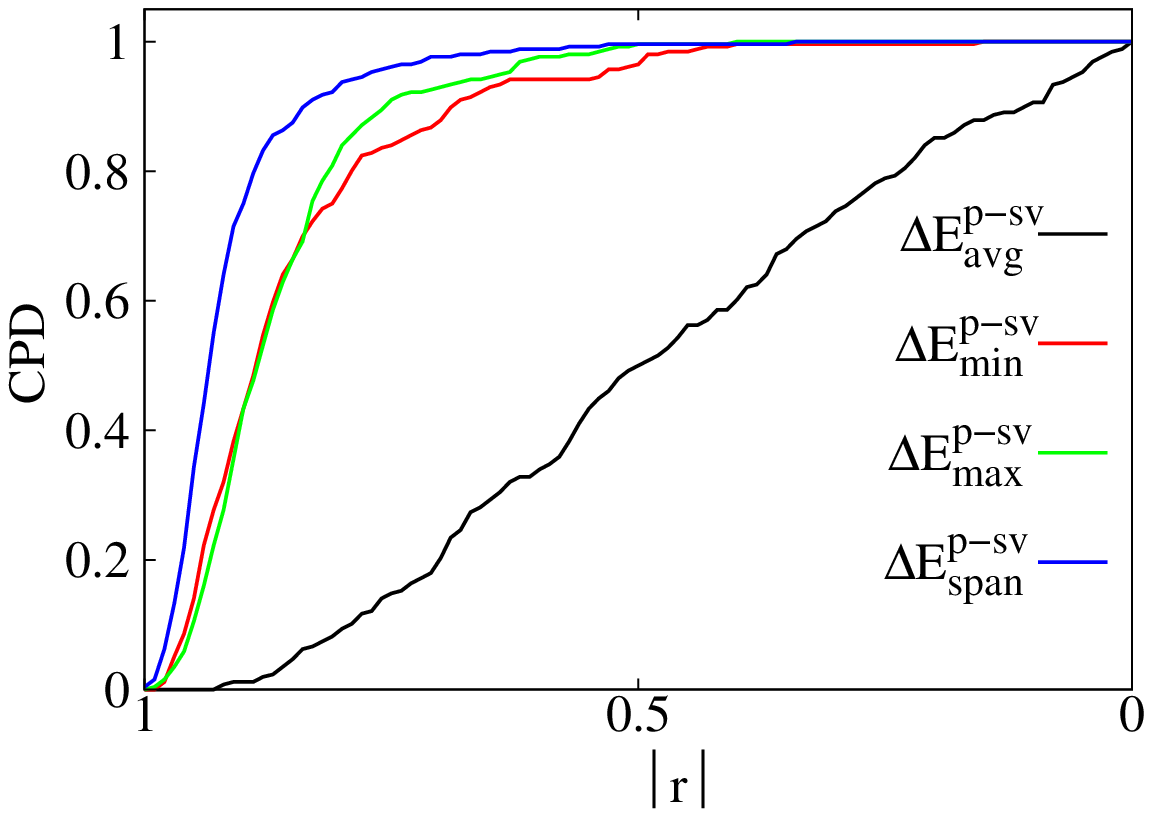}}\\
\subfloat[]{\includegraphics[width=1.5in]{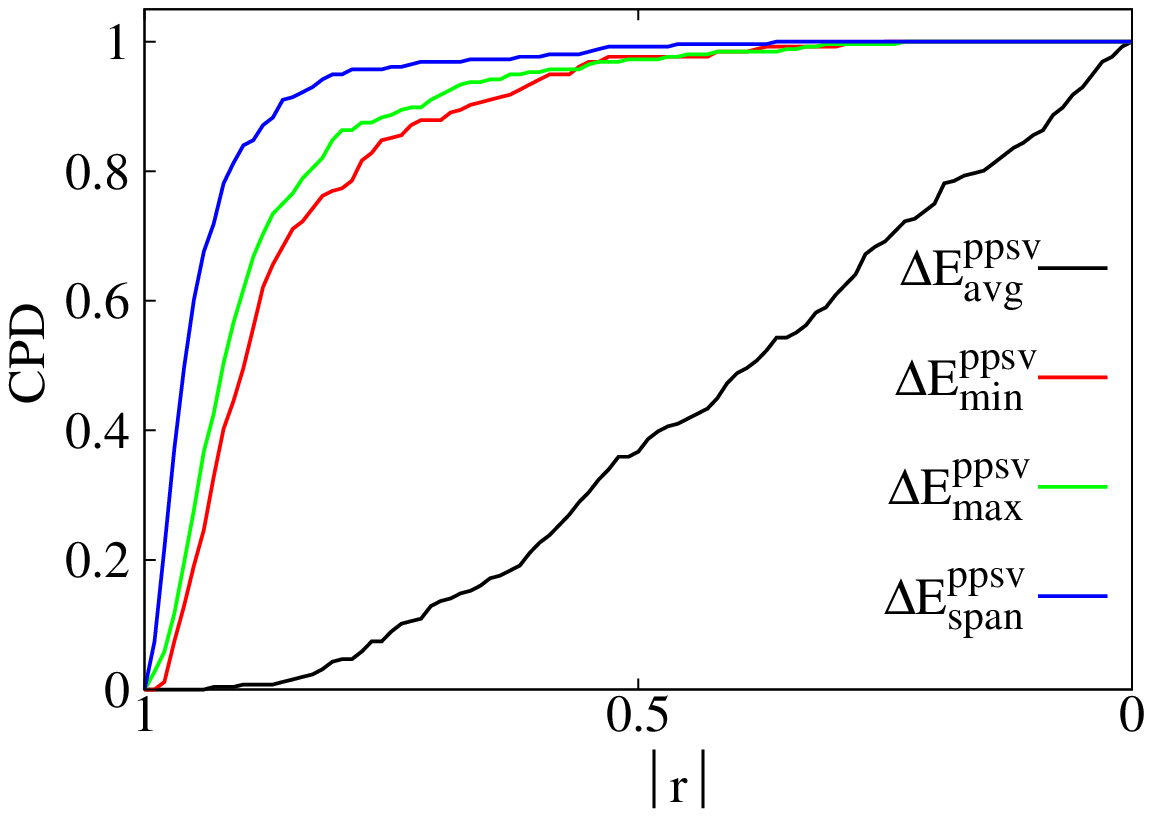}}
\subfloat[]{\includegraphics[width=1.5in]{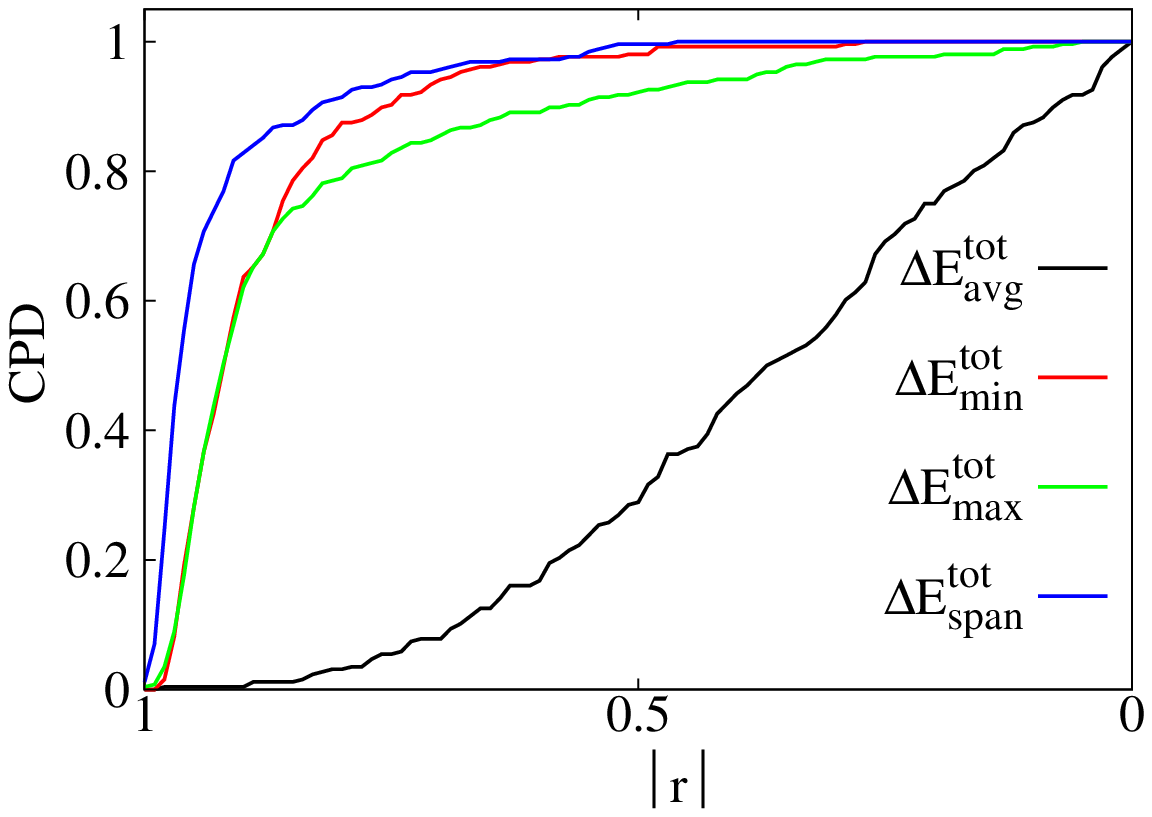}}
\caption{The cumulated probability density (CPD) of linear correlation coefficients between a)$\Delta E^p_\$$, b)$\Delta E^{p\tn{-}sv}_\$$, c)$\Delta E^{ppsv}_\$$ and d)$\Delta E^{tot}_\$$ (`\$' here is a wild card for $min$, $max$, $span$ and $avg$) and $\Delta F$  as a function of decreasing $|r|$ from 1 to 0.}
\label{fig:CPDpppt}
\end{figure}

For protein self energy, $\Delta E^p_{min}$ correlate slightly weaker than $\Delta E^p_{max}$ for large $r$, $\Delta E^p_{min}$ correlate slightly stronger than $\Delta E^p_{max}$ for small $r$ (Fig. \ref{fig:CPDpppt}a). For protein solvent interaction energy, the opposite is true (Fig. \ref{fig:CPDpppt}b). For the sum of protein self energy and protein solvent interaction energy, $\Delta E^{ppsv}_{min}$ exhibits weaker correlation with $\Delta F$ than $\Delta E^{ppsv}_{max}$ in the whole range of $r$ (Fig. \ref{fig:CPDpppt}c). Since MaPE terms, as briefly discussed earlier, qualitatively correlate to entropy in corresponding potential energy range, these observations suggest that for different potential energy terms, the relative importance of entropy is different.

In simulation and design of complex molecular systems exemplified by proteins, two fundamental limitations are accuracy of interaction representations (quality of force fields for classical systems) and sampling of the statistically significant region in configurational space. Scientists have been relentlessly working on these two fronts for decades with great progresses. However, a more practical general principle, \emph{which is to only generate information that is essential and utilize it to its full potential}, has not attracted sufficient attention. In regular protein docking, folding and design with the minimum potential energy utilized as the free energy proxy, backbone conformational states are selected as target macrostates first, side chain packing is repetitively carried out for a predetermined number of cycles and the minimum potential energy found is recorded to approximate free energy, and other information generated is discarded completely. With the addition of a variable for storing maximum potential energy and a subtraction operation to calculate corresponding span of potential energy, the reliability of prediction may potentially be significantly and consistently improved (Fig.\ref{fig:CPDpppt}).

It is hard to imagine that HEWL is an extremely special protein that has this unique property, which is not true for all other proteins and other complex bimolecular systems. To investigate general utility of potential energy span as a free energy proxy for other proteins, similar analyses were conducted for bovine pancreatic trypsin inhibitor (see Fig. S3 and Fig. S4) and human sulfotransferase (Fig. S5 and Fig. S6). As observed for HEWL, $E^*_{span}$ were observed to correlate consistently better with $\Delta F$ than both $E^*_{min}$ and $E^*_{max}$ for these two proteins. Therefore, it is likely that the observation is of general importance for proteins and possibly for other complex biomolecular systems as well.    

\section*{\sffamily \Large CONCLUSIONS}
In summary, based on the observation that change of MaPE terms exhibit similar level of correlations with change of free energy to change of MiPE terms, we proposed and tested span of potential energy as an approximate free energy proxy based on extensive MD trajectories of HEWL, and found that span of potential energy terms perform consistently and significantly better than corresponding minimum potential energy terms. Similar results were observed for two other different proteins. Therefore, it is likely that the superiority of span of potential energy as a free energy proxy to be of general importance for proteins and possibly other complex biomolecular systems. It is important to note that additional computational cost for obtaining span of potential energy is negligible. 

\subsection*{\sffamily \large ACKNOWLEDGMENTS}
This research was supported by National Natural Science Foundation of China under grant number 31270758, and by the Research fund for the doctoral program of higher education under grant number 20120061110019.


\clearpage


\bibliography{FreeEnergy3,Total}   
%
%
%


\clearpage

\captionsetup[subfigure]{labelformat=simple,position=bottom}



\clearpage


\end{document}